\documentclass[a4paper,11pt]{article}
\pdfoutput=1 % if your are submitting a pdflatex (i.e. if you have
\usepackage{jheppub} % for details on the use of the package, please
\usepackage[T1]{fontenc} % if needed
\usepackage{orcidlink}  % Orcid IDs
\usepackage[font=small,labelfont=bf]{caption}
\usepackage{float}
\usepackage{comment}

\title{\boldmath Light vector bosons and the weak mixing angle in the light of future germanium-based reactor CE$\nu$NS experiments}

\author[]{Manfred Lindner\,\orcidlink{0000-0002-3704-6016},}
\author[1]{Thomas Rink\note{Corresponding author.}\,\orcidlink{0000-0002-9293-1106}}
\author[]{and Manibrata Sen\,\orcidlink{0000-0001-7948-4332}}

\affiliation[]{Max-Planck-Institut f\"ur Kernphysik,\\
Saupfercheckweg 1, 69117 Heidelberg, Germany}

\emailAdd{manfred.lindner@mpi-hd.mpg.de}
\emailAdd{thomas.rink@mpi-hd.mpg.de}
\emailAdd{manibrata.sen@mpi-hd.mpg.de}

\abstract{
In this work, the sensitivity of future germanium-based reactor neutrino experiments to the weak mixing angle $\sin^{2}\theta_{W}$, and to the presence of new light vector bosons is investigated.
By taking into account key experimental features with their uncertainties and the application of a data-driven and state-of-the-art reactor antineutrino spectrum, the impact of detection threshold and experimental exposure is assessed in detail for an experiment relying on germanium semiconductor detectors.
With the established analysis framework, the precision on the Weinberg angle, and capability of probing the parameter space of a universally coupled mediator model, as well as a U(1)$_{\rm B-L}$-symmetric model are quantified.
Our investigation finds the next-generation of germanium-based reactor neutrino experiments in good shape to determine the Weinberg angle $\sin^{2}\theta_{W}$ with $<10$\% precision using the low-energetic neutrino channel of CE$\nu$NS.
In addition, the current limits on new light vector bosons determined by reactor experiments can be lowered by about an order of magnitude via the combination of both CE$\nu$NS and E$\nu$eS. 
Consequently, our findings provide strong phenomenological support for future experimental endeavours close to a reactor site. 
}

\begin{document} 
\maketitle
\flushbottom

%%%%%%%%%%%%%%%%%%%%%%%%%%%%%%%%%%%%%%%%%%%%%%%%%%%%%%%%%%%%%%%%%%%%%%%%%%%%%%%%%%%%%%%%%%%%%%%%%%%%

\section{Introduction}\label{sec:intro}

%%%%%%%%%%%%%%%%%%%%%%%%%%%%%%%%%%%%%%%%%%%%%%%%%%%%%%%%%%%%%%%%%%%%%%%%%%%%%%%%%%%%%%%%%%%%%%%%%%%%

With the first observations of coherent elastic neutrino-nucleus scattering (CE$\nu$NS)~\cite{Freedman:1973yd,Tubbs:1975jx,Freedman:1977xn} by the \textsc{Coherent} collaboration~\cite{COHERENT:2017ipa,COHERENT:2020iec} the pathway towards precision neutrino physics with small-scale detectors has been opened and promises a bright future with many opportunities for physics within the Standard Model (SM) and beyond (BSM).
Thus, it is not surprising that already these first measurements have been used for many SM and BSM investigations, targeting, for example, the weak mixing angle at low momentum transfers~\cite{Cadeddu:2018izq,Cadeddu:2020lky,AristizabalSierra:2022axl,Majumdar:2022nby,DeRomeri:2022twg}, the neutron radius~\cite{Cadeddu:2019eta,Papoulias:2019lfi,Cadeddu:2021ijh,DeRomeri:2022twg} in the context of the SM as well as electromagnetic properties of the neutrino~\cite{Papoulias:2017qdn,Cadeddu:2018dux,Khan:2019cvi}, non-standard neutrino interactions (NSI)~\cite{Lindner:2016wff,Papoulias:2017qdn,Denton:2018xmq,Khan:2019cvi,Coloma:2019mbs,Denton:2020hop,Denton:2022nol}, light mediators~\cite{Dent:2016wcr,Farzan:2018gtr,Miranda:2020zji,Cadeddu:2020nbr,Dutta:2022fdt}, new or sterile fermions~\cite{Brdar:2018qqj,Chang:2020jwl,Miranda:2020syh,Candela:2023rvt}, dark matter~\cite{Ge:2017mcq,COHERENT:2019kwz} and even axion-like particles~\cite{AristizabalSierra:2020rom} in the context of BSM physics. 

In the \textsc{Coherent} experiments, (anti-)neutrinos from pion-decays-at-rest ($\pi$DAR) were detected in multiple flavours: $\nu_{e}$, $\nu_{\mu}$, $\bar{\nu}_{\mu}$.   
The relatively high neutrino energies up to $\sim50$\,MeV and timing information about the incident neutrinos were among the key factors which enabled a successful CE$\nu$NS detection.
Additional measurements are planned with further target materials such as argon, germanium (Ge), sodium iodide and heavy water~\cite{Akimov:2022oyb}. 

\noindent With the aim of detecting this process at lower (antineutrino) energies ($E_{\nu}\lesssim10$~MeV), several experimental attempts have emerged close to nuclear reactors: \textsc{Connie}~\cite{CONNIE:2019swq,CONNIE:2024pwt}, \textsc{Conus}~\cite{CONUS:2020skt}, \textsc{Mi$\nu$er}~\cite{Ang:2021mdp}, \textsc{Ncc-1701}~\cite{Colaresi:2022obx}, \textsc{Neon}~\cite{Choi:2020gkm}, \textsc{Nucleus}~\cite{NUCLEUS:2019igx}, $\nu$GEN~\cite{nGeN:2022uje}, \textsc{Red-100}~\cite{Akimov:2022xvr}, \textsc{Ricochet}~\cite{Ricochet:2021rjo}, \textsc{Texono}~\cite{TEXONO:2020vnv}.   
These approaches try to exploit the naturally high, single neutrino flavour emission of about $\sim 10^{20}\, \bar{\nu}_{e}$/GW/s close to a reactor core and probe CE$\nu$NS with the full spectrum of modern detection technologies.
Although the intrinsically lower nuclear recoils from reactor antineutrinos render experimental conditions even more demanding, a strong CE$\nu$NS signal from a reactor site will provide a reference signal for future combined investigations of CE$\nu$NS data sets.
The lower momentum transfer in CE$\nu$NS of reactor antineutrinos allows us to measure the weak mixing angle at even lower energies~\cite{Kumar:2013yoa, ParticleDataGroup:2022pth,AristizabalSierra:2022axl}, and study new light degrees of freedom~\cite{Cerdeno:2016sfi,Papoulias:2017qdn,Papoulias:2019txv,Khan:2019cvi,Cadeddu:2020nbr,Miranda:2020tif,Miranda:2020zji,Bertuzzo:2021opb,Fernandez-Moroni:2021nap,DeRomeri:2022twg,AtzoriCorona:2022moj,Li:2022jfl,Coloma:2022avw}.
In this sense, both $\pi$DAR and reactor approaches effectively complement each other. Similarly, bounds on such light mediators can arise from the measurement of solar neutrinos~\cite{Demirci:2023tui}, which in turn affect the anticipated Migdal ionisation signals~\cite{Herrera:2023xun}.
Such constraints may additionally be accompanied by findings from cosmological observations~\cite{Sabti:2019mhn,Li:2023puz,Ghosh:2023ilw}.

The weak mixing angle $(\theta_W)$, also known as the Weinberg angle, is an important prediction of the SM. It relates the masses of the weak gauge bosons and forms an integral component of the electroweak precision observable tests. It can also be measured through a measurement of the SU$(2)_{\rm L}$ and the U$(1)_{\rm Y}$ couplings. As a result, a precise measurement of $\theta_W$ is required to test the validity of the SM. 

Naturally, by virtue of its definition, quantum corrections are relevant when determining $\theta_W$, and hence its value depends on the scale at which the measurements are made. This scale dependence has been confirmed by a number of different experiments at different scales (see \cite{ParticleDataGroup:2022pth} for a detailed review). The most precise measurements of the weak mixing angle come from the LEP and the SLC -both $e^+-e^-$ linear colliders- and use data collected at the $Z$-resonance to report $\sin^2\theta_W=0.23153\pm 0.00016$~\cite{ALEPH:2005ab}. 
Neutrino experiments have also joined the quest, with the most precise measurement of $\theta_W$ from neutrino scattering coming from the NuTEV experiment~\cite{NuTeV:2001whx}. Using deep-inelastic scattering of neutrinos and antineutrinos, NuTEV measured $\sin^2\theta_W=0.2407\pm 0.0016$ at an average energy scale $\mu=4.6\,{\rm GeV}$ - a result deviating from the LEP measurements by $3.6\sigma$. This makes it crucial for other neutrino scattering experiments to probe this discrepancy and check whether this is an effect of unaccounted (nuclear) physics or a signal of BSM physics~\cite{deGouvea:2019wav}. With the advent of CE$\nu$NS, it has become possible to measure $\theta_W$ at lower energy scales~\cite{Canas:2018rng,Miranda:2020tif,Fernandez-Moroni:2020yyl,DeRomeri:2022twg,AristizabalSierra:2022axl}. This allows us to test the running of $\theta_W$ at lower energies, thereby offering an important test of the quantum corrections affecting SM measurements.

Clearly, the measurement of $\theta_W$ is dependent on the underlying interaction channels for electrons and quarks. If hitherto unknown interactions are present for the SM particles, our inference of $\theta_W$ can change. This is particularly interesting if the new physics consists of light degrees of freedom since these can contribute to the kinematic regions away from the $Z$ boson pole. In fact, current measurements of $\theta_W$ are not as precise at low momentum transfers as they are at the $Z$-pole. This allows one to use CE$\nu$NS to test the interplay between $\theta_W$ and various simple well-motivated extensions of the SM. For example, some of the effects that could easily modify $\theta_W$ include the neutrino charge radius, a modified neutrino vector coupling, and the introduction of light new mediators, among others. 

These are the major issues we want to address in this work using future reactor antineutrino experiments. Further, Ge-based reactor experiments currently lead the CE$\nu$NS investigations at reactor sites~\cite{CONUS:2020skt,Colaresi:2022obx,nGeN:2022uje}, exhibit enormous scaling potential and feature ongoing improvements in detection sensitivity~\cite{Bonet:2023kob}. With its publications of several constraints on BSM physics~\cite{CONUS:2021dwh,CONUS:2022qbb}, we focus on the \textsc{Conus} experiment as a benchmark for the investigations of this work. Using different realisations of the experiment, we demonstrate the sensitivity to measuring the weak mixing angle within the SM at low momentum transfers of $\mathcal{O}(10)\,{\rm MeV}$. 
We obtain that the relative uncertainty of the mixing angle can improve to $\lesssim 10\%$ for feasible realisations of these experiments. Furthermore, we also test the potential of these experiments to text simple U(1) extensions of the SM involving a dark $Z'$ gauge boson. We consider two scenarios : (1) a universally coupled $Z'$, and (2) $Z'$ charged under a U(1)$_{\rm B-L}$. We find that, depending on the experimental considerations, the bounds on the couplings of these $Z'$ can be stronger by an order of magnitude than the current ones. This underlines the potential of high-purity Ge detectors for CE$\nu$NS experiments.

This work is structured in the following way:
Section~\ref{sec:channels} introduces the interactions under study and discusses their individual characteristics.
A generic Ge-based reactor experiment is introduced in section~\ref{sec:exp_spec}, together with the analysis framework we utilise for our investigations.
The main results are presented in terms of sensitivity estimates on the weak mixing angle as well as on light vector bosons in section~\ref{sec:results}.
Finally, we conclude with a summary and an outlook in section~\ref{sec:conclusions}.

%%%%%%%%%%%%%%%%%%%%%%%%%%%%%%%%%%%%%%%%%%%%%%%%%%%%%%%%%%%%%%%%%%%%%%%%%%%%%%%%%%%%%%%%%%%%%%%%%%%%

\section{Neutrino interactions in germanium detectors close to a nuclear reactor}\label{sec:channels}

In what follows we introduce the SM neutrino interaction channels relevant for the analysis of this work, together with potential BSM modifications: elastic neutrino-electron scattering (E$\nu$eS) and coherent elastic neutrino-nucleus scattering (CE$\nu$NS). 
Both are, in principle, sensitive to the weak mixing angle and new light vector bosons, for which we assume typical benchmark models.

%%%%%%%%%%%%%%%%%%%%%%%%%%%%%%%%%%%%%%%%%%%%%%%%%%%%%%%%%%%%%%%%%%%%%%%%%%%%%%%%%%%%%%%%%%%%%%%%%%%%

\subsection{Elastic neutrino scatterings in the Standard Model}

Antineutrinos emitted by a nuclear reactor can interact in two different ways within a semiconductor detector, either via scattering off the electrons or the nuclei of the underlying target material.
Elastic scatterings of (anti-)neutrinos off electrons are mediated by the exchange of electroweak gauge bosons and can be expressed in the following compact differential cross section~\cite{Vogel:1989iv}:
\begin{align}\label{eq:cross_section_sm_nu_e}
\frac{\mathrm{d}\sigma}{\mathrm{d}T_{e}} (T_{e}, E_{\nu}) = \frac{G_{F}^{2} m_{e}}{2\pi} \Big[ \left(g_{V} + g_{A}\right)^{2} + \left(g_{V} - g_{A}\right)^{2}\left(1-\frac{T_{e}}{E_{\nu}} \right)^{2} + \left(g_{A}^{2} - g_{V}^{2}\right)\frac{m_{e} T_{e}}{E_{\nu}^{2}} \Big]\, ,
\end{align}
where $T_{e}$ represents the electron recoil energy after the interaction, $G_{F}$ the Fermi constant, $m_{e}$ the electron mass and $E_{\nu}$ the energy of the incident (anti-)neutrino.
The axial-vector and vector couplings $g_{A,V}$ for the underlying currents depend on the neutrino flavour and are defined as 
\begin{align}\label{eq:electroweak_couplings_electron_scattering}
    g_{V}&=
    \begin{cases} \frac{1}{2} + 2\sin^{2}\theta_{W}\, , & \text{for } \nu_{e}\, , \\ 
    -\frac{1}{2} + 2\sin^{2}\theta_{W}\, , & \text{for } \nu_{\mu},\nu_{\tau}\, ,
    \end{cases}
    &g_{A}= 
    \begin{cases} \frac{1}{2}\, , & \text{for } \nu_{e}\, , \\ 
    -\frac{1}{2}\, , & \text{for } \nu_{\mu},\nu_{\tau}\, ,
    \end{cases}
\end{align}
while in the case of antineutrino scattering, the axial-vector couplings change their sign, i.e.\ $g_{A}\rightarrow -g_{A}$.

For coherent elastic neutrino scattering off a nucleus, the neutrino energy must not be too large such that the mediator's wavelength extends over the whole nucleus. This coherence condition is usually given by $qR\sim1$, where $q$ is the reaction's momentum transfer and $R$ is the nucleus radius. 
This can be translated into a condition on the energy of the incident neutrino: for energies below $E_{\nu}\sim$19\,MeV, which is the case for reactor antineutrinos, coherent interactions of neutrinos scattering off a Ge nucleus can be assumed. This corresponds to a maximum nuclear recoil energy of $T_{A}\sim11$\,keV.

The differential CE$\nu$NS cross section is given by~\cite{Freedman:1973yd}
\begin{equation}\label{eq:cross_section_sm_cenns}
\begin{aligned}
    \frac{d\sigma}{dT_{A}}(T_{A}, E_{\nu}) & = \frac{G_{F}^{2} m_{A}}{4 \pi} Q_{W}^{2} \left(1-
\frac{m_{A} T_{A}}{2 E^{2}_{\nu}}\right) |F(T_{A})|^{2}\, , \\\text{with } Q_{W} &=\left[(1-4\sin^{2}\theta_{W})Z - N\right]\, .
\end{aligned}
\end{equation}
Here, $m_{A}$ is the nucleus mass and $Q_{W}$ is the weak nuclear charge, which is defined by the number of protons $Z$ and neutrons $N$, respectively.
For the nuclear form factor, the Helm parameterisation is used
\begin{align}
    F(T_{A}) = \frac{3j_{1}(q(T_{A}) R_{1})}{q(T_{A}) R_{1}}\exp\left[-\frac{(q(T_{A})s)^{2}}{2}\right]\, ,
\end{align}
with the spherical Bessel function $j_{1}$, the momentum transfer $q^{2}=2m_{A}T_{A}$, the nuclear skin thickness $s\simeq1$~fm, $R_{1}=\sqrt{R^{2} - 5s^{2}}$ and $R\simeq 1.2 A^{1/3}$~fm~\cite{Helm:1956zz,Engel:1991wq}.
Due to the smaller momentum transfer at a reactor site, the impact of the nuclear form factor is negligible, i.e.\ $F\rightarrow1$~\cite{CONUS:2021dwh}.
However, with increasing precision on the SM signal - as investigated in this work - this quantity becomes more relevant.

The target recoil energies $T_{x}$ for $x=\{e, A\}$ depend on the scattering angle $\theta$ (lab frame) and is given by
\begin{align}\label{eq:nuclear_recoil_energy}
T_{x}=\frac{2m_{x} E^{2}_{\nu} \cos^{2}\theta }{(m_{x} + E_{\nu})^{2} - E^{2}_{\nu} \cos^{2}\theta } \xrightarrow{\theta\rightarrow 0} \frac{2 E^{2}_{\nu}}{m_{x}+ 2 E_{\nu}}\, ,
\end{align}
where the last step defines the maximal nuclear recoil $T_{x}^{\mathrm{max}}$.

While measuring E$\nu$eS is a rather straightforward procedure within a semiconductor detector (recoiling electrons can directly be read out as charge signals), the read-out of nuclear recoil energies left by CE$\nu$NS is more involved for this detection technology that only detects charge/ionisation signals.
In the conversion process from nuclear to ionisation energy after a CE$\nu$NS event, so-called signal quenching, roughly 80\% of the initial energy is lost in dissipative processes and, thus, not accessible.
The ratio between ionisation energy $E_{I}$ and nuclear recoil energy $T_{A}$ is called the quenching factor and an energy-dependent quantity.
Here, the conventional Lindhard model is used to quantify signal quenching in Ge detectors~\cite{Lindhard:1961zz,Bonhomme:2022lcz}.\footnote{Currently, deviations from the Lindhard descriptions have been reported at sub-keV energy~\cite{Collar:2021fcl}. Since proposed physical explanations for an increased quenching effect have been excluded~\cite{AtzoriCorona:2023ais}, the authors only apply the conventionally used quenching model of Lindhard. However, we would like to emphasise the importance of quenching for reliable CE$\nu$NS predictions and turn to the experimental community to quickly solve this problem.} 

%%%%%%%%%%%%%%%%%%%%%%%%%%%%%%%%%%%%%%%%%%%%%%%%%%%%%%%%%%%%%%%%%%%%%%%%%%%%%%%%%%%%%%%%%%%%%%%%%%%%

\subsection{Modifications due to New Physics}

The interaction channels presented above can be subject to modification of BSM physics if new interactions with electrons and quarks are introduced.
Here, we assume the existence of an additional $Z$-like gauge boson with interactions given by the following Lagrangian:
\begin{align}\label{eq:lagrangian_light_vector}
\mathcal{L}_{Z'} = Z'_{\mu} \left( g^{\nu \mathrm{V}}_{Z'} \bar{\nu}_{L} \gamma^{\mu} \nu_{L}
+  g^{e \mathrm{V}}_{Z'} \bar{e} \gamma^{\mu} e  + g^{q \mathrm{V}}_{Z'} \bar{q} \gamma^{\mu} q \right) 
+ \frac{1}{2} m^{2}_{Z'} Z'_{\mu}Z'^{\mu}\, ,
\end{align}
with the new vector couplings $g^{x\mathrm{V}}_{Z'}$ for $x\in\{\nu,e,q\}$, the boson mass $m_{Z'}$ and $q\in\{u,d\}$ being the first generation of quarks.
For the time being, we leave aside any aspects of model-building and only assume the existence of an appropriate mass mechanism for the new light gauge boson.
In this sense, we follow the framework of simplified models usually used in collider studies~\cite{Salvioni:2009mt,Abercrombie:2015wmb,Cerdeno:2016sfi,Arcadi:2017kky,Morgante:2018tiq}.
Its coherent interaction with nuclei contributes to another weak nuclear charge that is given by
\begin{align}\label{eq:add_weak_charge}
    Q_{Z'} = \left(2g^{u \mathrm{V}}_{Z'} + g^{d \mathrm{V}}_{Z'}\right)Z + \left( g^{u \mathrm{V}}_{Z'} + 2 g^{d \mathrm{V}}_{Z'}\right) N\, .
\end{align}
As a consequence, the weak nuclear charge appearing in the conventional CE$\nu$NS cross-section, cf.~eq.~\eqref{eq:cross_section_sm_cenns}, is modified by the presence of the new vector couplings: $Q_{W}\rightarrow Q_{W} + Q_{Z'}$.

The same applies to the $Z$-boson contribution of E$\nu$eS, but with the difference that eq.~\eqref{eq:cross_section_sm_nu_e} also contains a contribution due to the exchange of $W$ boson.
The modified E$\nu$eS cross section due to the exchange of an additional $Z$-like boson is given by~\cite{Cerdeno:2016sfi}
\begin{align}\label{eq:cross_section_ligth_vector_nu_e}
\left(\frac{d\sigma}{dT_{e}}\right)_{\mathrm{E\nu eS\,+\,Z'} }=\left(\frac{d\sigma}{dT_{e}}\right)_{\mathrm{E\nu eS}} 
+ \frac{\sqrt{2} G_{F} m_{e} g_{V} g_{Z'}^{\nu \mathrm{V}} g_{Z'}^{e \mathrm{V}} }{\pi (2m_{e}T_{e}  + m_{Z'}^{2})} 
+ \frac{m_{e} (g_{Z'}^{\nu \mathrm{V}} g_{Z'}^{e \mathrm{V}})^{2} }{2\pi (2m_{e}T_{e}  + m_{Z'}^{2})^{2}}\, ,
\end{align}
with the usual SM vector coupling $g_{V}$ defined in eq.~\eqref{eq:electroweak_couplings_electron_scattering} as well as the new $Z$ couplings $g^{\nu V}_{Z'}$ and $g^{eV}_{Z'}$ to neutrinos and electrons, respectively. 

In the end, we would like to mention that the explicit appearance of the $Z'$ propagator term in both, E$\nu$eS and CE$\nu$NS, cross sections divides the model's parameter space in two different kinematic regions: $q\ll m_{Z'}^{2}$ and $q\gg m_{Z'}^{2}$.
In the latter case, the dependence on $m_{Z'}$ is negligible and limits become independent 
of the new boson mass. In the following, we introduce the models that are investigated in the course of this work:

%%%%%%%%%%%%%%%%%%%%%%%%%%%%%%%%%%%%%%%%%%%%%%%%%%%%%%%%%%%%%%%%%%%%%%%%%%%%%%%%%%%%%%%%%%%%%%%%%%%%

\paragraph{Universal coupling}
A straightforward way to simplify the parameter space of the introduced U(1) extension of the SM is to assume equal coupling to all (or at least the first generation of) SM fermions, i.e.\ $g_{Z'} \equiv g^{\nu \mathrm{V}}_{Z'}=g^{e \mathrm{V}}_{Z'}=g^{u \mathrm{V}}_{Z'}=g^{d \mathrm{V}}_{Z'}$.
In doing so, the relevant parameter space is shrunk to only two parameters: the universal coupling $g_{Z'}$ and the boson mass $m_{Z'}$.
Due to this charge assignment, the additional nuclear charge of eq.~\eqref{eq:add_weak_charge} simplifies further $Q_{Z'} \rightarrow 3\, g_{Z'} \left( Z+N \right)$ such that the combined CE$\nu$NS cross section is given by~\cite{Cadeddu:2020nbr}
\begin{align}\label{eq:cross_section_vector_universal}
    \frac{d\sigma}{dT_{A}}(T_{A}, E_{\nu}) & = \frac{G_{F}^{2}m_{A}}{\pi} \bigg[ Q_{W} + \frac{3 g_{Z'}^{2}}{\sqrt{2}G_{F}} \frac{Z+N}{2m_{A}T_{A} + m_{Z'}^{2}}\bigg]^{2} \left(1-
\frac{m_{A} T_{A}}{2 E^{2}_{\nu}}\right) |F(T_{A})|^{2}\, .
\end{align}
The quadratic dependency on the weak nuclear charge in eq.~\eqref{eq:cross_section_vector_universal} and the fact that $Q_{W}<0$ for practical purposes, induces a degeneracy in the parameter space.  
In case of $\big[Q_{W} + Q_{Z'}]\simeq \pm Q_{W}$, the experimental signature appear as SM-like.
This leads to a situation in which the light vector mediator contribution cannot be distinguished from SM CE$\nu$NS and an exclusion demands targets with varying proton and neutron numbers as well as different sensitivity to the nuclear recoil energies left in the detector. 

%%%%%%%%%%%%%%%%%%%%%%%%%%%%%%%%%%%%%%%%%%%%%%%%%%%%%%%%%%%%%%%%%%%%%%%%%%%%%%%%%%%%%%%%%%%%%%%%%%%%

\paragraph{The benchmark model U(1)$_{\rm B-L}$}

One of the simplest and most widely studied extensions of the SM is an additional vector boson with coupling to a particle's B$-$L charge.
Besides a quite natural definition of charge, i.e.\ $g_{Z'} \equiv g^{\nu \mathrm{V}}_{Z'}=g^{e \mathrm{V}}_{Z'}=-g^{u \mathrm{V}}_{Z'}/3=-g^{d \mathrm{V}}_{Z'}/3$, this model arises from a variety of more complete SM extensions and is by definition free of any gauge anomalies.
Thus, it appears as a valid representation of nature with an economic parameter space spanned only by the boson mass $m_{Z'}$ and the absolute coupling strength $g_{Z'}$.
The resulting modified CE$\nu$NS cross-section is then given by
\begin{align}\label{eq:cross_section_vector_b_l}
    \frac{d\sigma}{dT_{A}}(T_{A}, E_{\nu}) & = \frac{G_{F}^{2}m_{A}}{\pi} \bigg[ Q_{W} - \frac{ g_{Z'}^{2}}{\sqrt{2}G_{F}} \frac{Z+N}{2m_{A}T_{A} + m_{Z'}^{2}}\bigg]^{2} \left(1-
\frac{m_{A} T_{A}}{2 E^{2}_{\nu}}\right) |F(T_{A})|^{2}\, .
\end{align}
In this model, the parameter space is non-degenerate such that conclusive exclusion limits can be deduced. Since the charge assignment in both models is the same for neutrinos and electrons, both lead to the same modification of E$\nu$eS. Therefore, we expect only additional events compared to the SM case for E$\nu$eS.

%%%%%%%%%%%%%%%%%%%%%%%%%%%%%%%%%%%%%%%%%%%%%%%%%%%%%%%%%%%%%%%%%%%%%%%%%%%%%%%%%%%%%%%%%%%%%%%%%%%%

\begin{figure}
  \begin{minipage}{.49\linewidth}
    \centering
    \includegraphics[width=\textwidth]{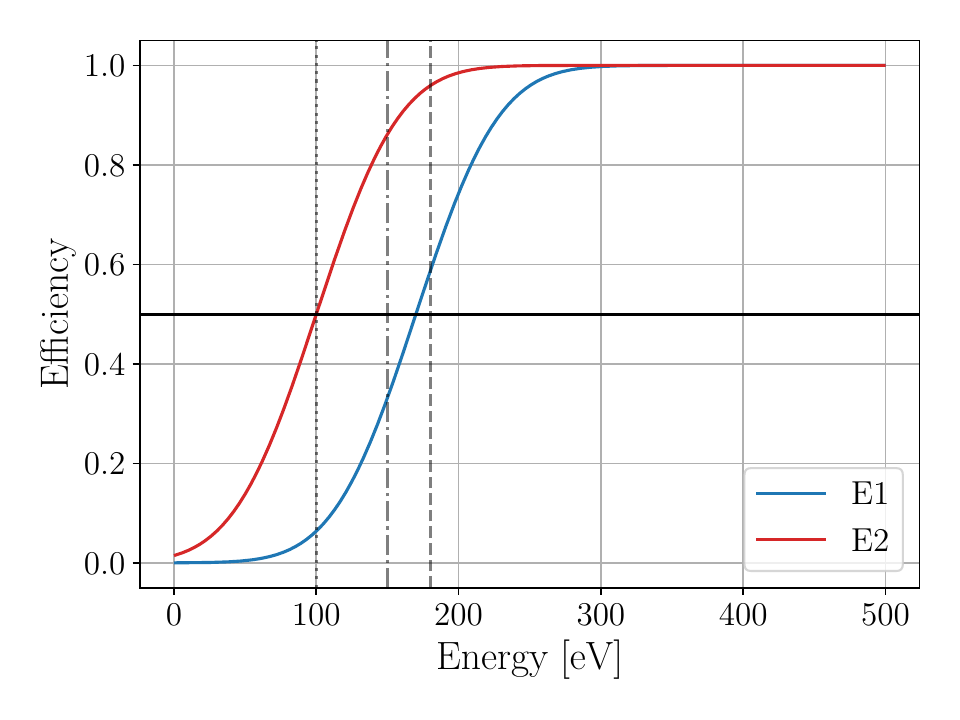}
	\captionof{figure}{
	    Detector (trigger) efficiencies applied throughout this work. The blue curve - as achieved in ref.~\cite{Bonhomme:2022lcz} - is used for the conservative and the expected threshold scenario, while the red curve corresponds to our optimistic case. Vertical lines indicate the three threshold benchmarks, cf.~table~\ref{tab:experimental_configurations}. 
	}
	\label{fig:trigger_efficiency}
  \end{minipage}\hfill
  \begin{minipage}{.49\linewidth}
    \centering
    \vspace{1cm}
    \begin{tabular}{ c || c }
      det. thresholds [eV] & exp. exposure  \\
      \hline
       180 (E1) & 5 kg$\cdot$yr \\
       150 (E1) & 50 kg$\cdot$yr \\
       100 (E2) & 500 kg$\cdot$yr \\
    \end{tabular}
    \vspace{1.5cm}
    \captionof{table}{Overview of assumed detector thresholds and experimental exposures. The applied detector efficiencies are given in brackets and shown in figure~\ref{fig:trigger_efficiency}. All combinations of these values are considered in the analysis below.}
    \label{tab:experimental_configurations}
  \end{minipage}
\end{figure}

%%%%%%%%%%%%%%%%%%%%%%%%%%%%%%%%%%%%%%%%%%%%%%%%%%%%%%%%%%%%%%%%%%%%%%%%%%%%%%%%%%%%%%%%%%%%%%%%%%%%

\section{Experimental set-up and investigation strategy}\label{sec:exp_spec}

After clarifying the phenomenological foundations of this work, we define the experimental framework of this analysis and the methodology of our sensitivity estimation.
Since Ge detectors are currently leading CE$\nu$NS investigations at reactor sites~\cite{CONUS:2020skt,Colaresi:2022obx,nGeN:2022uje}, the main focus of this work are future capabilities of Ge-based experiments. We point out that the assumptions made in the following strongly depend on the chosen detection technology and differ from other experiments, e.g. with respect to the achievable background level, the detection efficiency and the threshold energy. In particular, the Ge technology assumed here is not able to distinguish between electron and nuclear recoil events.

%%%%%%%%%%%%%%%%%%%%%%%%%%%%%%%%%%%%%%%%%%%%%%%%%%%%%%%%%%%%%%%%%%%%%%%%%%%%%%%%%%%%%%%%%%%%%%%%%%%%

\subsection{Experimental set-up under study}

As the strongest artificial (anti-)neutrino source on Earth, we select a commercial nuclear power plant with a thermal power of 3.5~GW, a value in the ballpark of currently running pressurised water reactors.
A distance of $\sim20$~m from the reactor core represents a good trade-off between potential reactor-correlated background events and the expected neutrino flux~\cite{Hakenmuller:2019ecb}.
With these assumptions a total antineutrino flux of $\phi\sim 1.5\cdot10^{13}\, \bar{\nu}_{e}$/cm$^{2}$/s is obtained.
For the reactor antineutrino emission spectrum, we use a recently proposed data-driven approach~\cite{DayaBay:2021dqj}.
For this, information from inverse beta decay (IBD) measurements is used to construct an antineutrino emission spectrum tailored to a given reactor fuel composition.
We assume a typical configuration for a pressurised water reactor, where the antineutrinos are mainly emitted in beta decay chains of four isotopes.
Thus, following fission fractions are used: ($^{235}$U, $^{238}$U, $^{239}$Pu, $^{241}$Pu) = (56.1, 7.6, 30.7, 5.6)\,\%.
The unfolded IBD spectra of ref.~\cite{DayaBay:2021dqj} (1.8-7~MeV), are combined with new measurements of the spectrum's high energy part (7-11~MeV)~\cite{DayaBay:2022eyy} and low-energy information from updated summation methods (0-1.8~MeV)~\cite{Estienne:2019ujo}.

As an experimental target, a Ge semiconductor detector with three benchmark exposures is assumed: 5\,kg$\cdot$yr, 50\,kg$\cdot$yr and 500\,kg$\cdot$yr.
Since commercial power plants exhibit short maintenance periods of $\sim1$ month/year, experiments close to reactors are usually limited by their background measurements, i.e.\ the reactor OFF periods.
We take this into account by using different run times for reactor ON and OFF periods and optimistically assume $t_{\mathrm{OFF}} =0.5\cdot t_{\mathrm{ON}}$.
For the detection threshold $E_{\mathrm{thr}}$, we follow current detector developments and use again three different values in our investigation:  180~eV (conservative), 150~eV (expected) and 100~eV (optimistic).
An overview of the assumed experimental parameters is given in table~\ref{tab:experimental_configurations}.
To describe signal quenching, i.e.\ the conversion of nuclear recoil energies to ionisation signal, for CE$\nu$NS in our detectors, we follow recent measurements and apply Lindhard's model~\cite{Lindhard:1961zz,Bonhomme:2022lcz} down to detection energies of 100~eV.
In doing so, we remain conservative since increased signal quenching for whatever reason also increases the expected signal counts and, thus, the experimental reach.
An energy-dependent experimental resolution is incorporated via convolutions with a Gaussian whose width is also determined by the chosen experimental detection threshold.
In reality, a heuristic relation between a detector's energy threshold and a pulser's full width at half maximum (FWHM), i.e.\ FWHM of an artificial pulse at zero detection energy, is observed:  $E_{\mathrm{thr}}\sim 3 \cdot$ FWHM$_{\mathrm{pulser}}$.
Thus, in the convolution that takes care of the detector resolution, the Gaussian width is given by the following expression:
\begin{align}\label{eq:detector_resolution}
    \sigma(E) = \sqrt{\left( \frac{E_{\rm thr}}{ 3\cdot2.355}\right)^{2} + \epsilon \cdot \mathcal{F} \cdot E}\, ,  
\end{align}
with the assumed detection threshold, $E_{\rm thr}$, the energy needed to create an energy-hole pair at 90K in a Ge semiconductor, $\epsilon=2.96$\,eV, and the Fano factor for Ge, given by $\mathcal{F}=0.11$. The first term represents the pulser resolution (FWHM) at zero energy.
To incorporate realistic detector efficiencies, we apply the (trigger) efficiency curves shown in figure~\ref{fig:trigger_efficiency}.\footnote{The curves are basically Gaussian CDFs with parameters chosen to fit realistic detection efficiencies at lowest energies~\cite{Bonhomme:2022lcz,Bonet:2023kob}.}
For the conservative and expected threshold case, the blue curve is applied~\cite{Bonhomme:2022lcz}, whereas the optimistic case uses a curve that exhibits 50\% detection efficiency down to the threshold of 100~eV.
The first is motivated by recent developments in Ge semiconductor detectors~\cite{Bonhomme:2022lcz}, whereas the red curve has been manually chosen to reach 50\% efficiency down to $E_{I}=100$\,eV.
Although recent studies show that additional background reduction techniques such as pulse shape discrimination by (charge) signal rise times are in principle applicable at such low energies~\cite{Bonet:2023kob}, their implementation with additional cuts will depend heavily on the individual experimental configuration and is thus beyond the scope of this work. 

A crucial ingredient of any CE$\nu$NS measurement is to prove that reactor-correlated background, especially neutrons, are under control~\cite{Hakenmuller:2019ecb}, which we assume to be the case.
The experimental background level depends strongly on the individual experimental set-up, in particular the applied detector and shield materials, and the overburden and radioactivity on site.
For more details on experimental background close to reactors, we refer the reader to ref.~\cite{Bonet:2021wjw}.
For this sensitivity estimate, we simply assume flat and constant background levels, which is at least justified far away from the detector's electronic noise and for keV-energies.
For the energy region $1~\mathrm{keV}< E_{I} \leq 10~\mathrm{keV}$, mainly important for E$\nu$eS, we assume a background level of $\sim 0.5$\,cnts/kg/d/keV.
For energies $E_{I}\leq1~\mathrm{keV}$ that are relevant for CE$\nu$NS investigations, the obtained background levels are usually higher and rise towards the detector threshold.
Thus, we assign it an overall background level of $\sim10$\,cnts/kg/d/keV, which was achieved by \textsc{Conus} on their first experimental runs~\cite{CONUS:2020skt}.
Further improvements in these values might be expected in the future.

%%%%%%%%%%%%%%%%%%%%%%%%%%%%%%%%%%%%%%%%%%%%%%%%%%%%%%%%%%%%%%%%%%%%%%%%%%%%%%%%%%%%%%%%%%%%%%%%%%%%

\subsection{Sensitivity estimation}

To determine the experimental sensitivity,  we closely follow the method used by the \textsc{Conus} collaboration and construct a likelihood function that fits reactor ON and OFF periods simultaneously, while experimental uncertainties are incorporated with Gaussian pull terms ~\cite{CONUS:2020skt,CONUS:2021dwh}.
Hence, the used (binned) likelihood function exhibits the form
\begin{align}
        -2\log\mathcal{L} = -2\log\mathcal{L}_{\mathrm{ON}} -2\log\mathcal{L}_{\mathrm{OFF}} + \sum_{i} \frac{(\theta_{i} - \bar{\theta}_{i})^2}{\sigma_{\theta_{i}}}\, ,
\end{align}
where $\theta_{i}$ are the incorporated nuisance parameters, $\bar{\theta}_{i}$ their best-fit values and $\sigma_{\theta_{i}}$ the corresponding uncertainties.
For this phenomenological analysis, we assume experimental quantities like detection efficiency, the detector's energy calibration and resolution as well as active volume to be sufficiently known such that corresponding uncertainties play a minor role and can be safely ignored.
The only experimental uncertainties incorporated in this analysis are attributed to signal quenching (Lindhard model with $\sim1$~\% uncertainty in its $k$ parameter, which reflects the quenching factor at $T_{A}\sim 1$\,keV~\cite{Bonhomme:2022lcz}) and an overall reactor antineutrino flux normalisation (3\%) originating from uncertainties on its thermal power and the underlying fission fractions.
Furthermore, the background contribution in both energy regions is assigned an unconstrained normalisation parameter $b$ in order to account for potential fluctuations.
Thus, the constructed analysis framework considers the following quantities:
\begin{itemize}
    \item weak mixing angle $\sin^{2}\theta_W$, vector boson coupling $g_{Z'}$ and mass $m_{Z'}$
    \item two background normalisation parameters: $b_{\mathrm{low}}$ ($E<1$\,keV) and $b_{\mathrm{high}}$ ($E>1$\,keV)
    \item two (pulled) nuisance parameters: Lindhard (quenching) parameter $k=0.162\pm0.004$ and reactor antineutrino flux $\phi \sim (1.50\pm0.04)\cdot10^{13}\, \bar{\nu}_{e}$/cm$^{2}$/s 
\end{itemize}
For the investigation of light vector bosons, we use knowledge from previous investigations of the Weinberg angle and include this via an additional Gaussian pull-term.
Here, we use a recent data-driven low-energy determination through combined data sets of atomic-parity violation investigations on caesium and \textsc{Coherent} CE$\nu$NS measurements, i.e.\ $\sin^{2}\theta_{W} = 0.2374\pm0.0020$~\cite{AtzoriCorona:2023ktl}. 

In the following, we estimate the future experimental sensitivity with two different methods: Asimov and Monte-Carlo (MC)-sampled data sets~\cite{Cowan:2010js,ParticleDataGroup:2022pth}. 
We apply a (profile) likelihood ratio test and deduce limits from a $\chi^{2}$-distributed test statistics.\footnote{For the investigation of the weak mixing angle, we assume a two-sided statistics test, since both downward and upward fluctuations might occur. The gauged U(1)$_{\rm B-L}$ model, only contributes additional signal events, thus, a one-sided statistics test is applied. For the case of the universally coupled model, the degeneracy of the parameter space also allows for a reduction of signal events for CE$\nu$NS, while E$\nu$eS can still be enhanced. This degeneracy region is indicated in the following results plots and limits on the smallest coupling strength are determined via one-sided test statistics.
}
While the first method directly yields exclusion limits that represent the full exploitation of statistical information, the second approach gives a distribution of potential limits, from which the median average is commonly defined as the expected experimental sensitivity. 

%%%%%%%%%%%%%%%%%%%%%%%%%%%%%%%%%%%%%%%%%%%%%%%%%%%%%%%%%%%%%%%%%%%%%%%%%%%%%%%%%%%%%%%%%%%%%%%%%%%%

\begin{figure}
    \centering
    \includegraphics[width=0.7\textwidth]{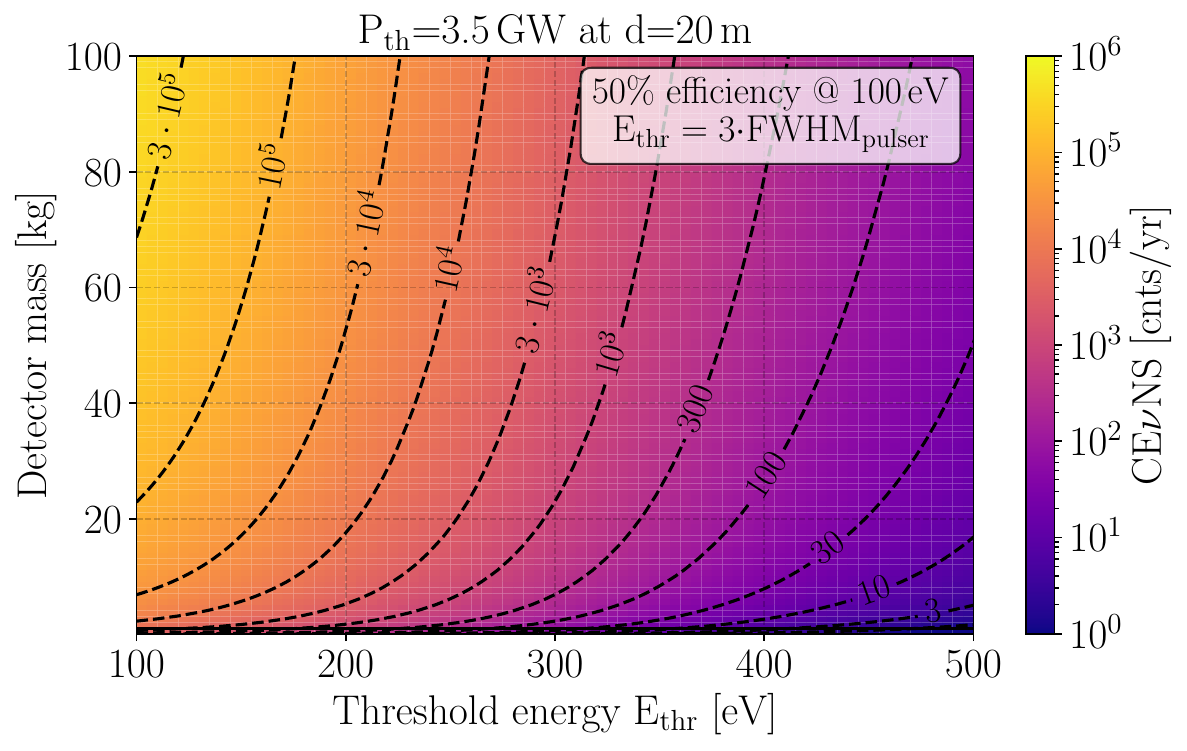}
    \caption{CE$\nu$NS signal expectation in dependence of target mass and detection threshold for the chosen experimental site close to a reactor. The optimistic detector efficiency of figure~\ref{fig:trigger_efficiency} has been chosen. The detector resolution is dynamically set in dependence on the assumed threshold, cf.~eq.~\eqref{eq:detector_resolution}.}
    \label{fig:signal_threshold_mass}
\end{figure}

%%%%%%%%%%%%%%%%%%%%%%%%%%%%%%%%%%%%%%%%%%%%%%%%%%%%%%%%%%%%%%%%%%%%%%%%%%%%%%%%%%%%%%%%%%%%%%%%%%%%

\section{Results}\label{sec:results}

Here, we highlight the effects of key experimental quantities like detection threshold and experimental exposure, and present the results of our sensitivity estimate on the weak mixing angle $\sin^{2}\theta_{W}$ and the parameter space of two light vector boson models.
The expected sensitivities are determined for the combination of three different detector thresholds (100~eV, 150~eV, 180~eV) and three potential experimental exposures (5~kg$\cdot$yr, 50~kg$\cdot$yr, 500~kg$\cdot$yr), cf.~table~\ref{tab:experimental_configurations}. 

%%%%%%%%%%%%%%%%%%%%%%%%%%%%%%%%%%%%%%%%%%%%%%%%%%%%%%%%%%%%%%%%%%%%%%%%%%%%%%%%%%%%%%%%%%%%%%%%%%%%

\subsection{Signal expectations in future germanium-based reactor experiments}

Before we present our results, we calculate the number of expected SM events to quantify the statistical reach we could expect for the chosen experimental configurations.
The SM E$\nu$eS is rather constant up to detection energies of 100~keV with an event rate of $\sim113$~cnts/kg/yr independent of the chosen detection threshold $E_{\mathrm{thr}}$.

Its strong sensitivity reach stems from the fact that the SM contribution is negligible compared to the assumed background contributions and that $\mathcal{O}(1)$ couplings of BSM models already contribute sizable event rates.
Due to the kinematic term in eq.~\eqref{eq:cross_section_sm_cenns} and the effect of signal quenching, CE$\nu$NS only extends up to energies of $\sim750$~eV, however with contributions that by far exceed the electron channel: (655, 988, 4481)~cnts/kg/yr for (180, 150, 100)~eV.\footnote{A reactor antineutrino with $E_{\nu}$=11\,MeV, leaves a maximal recoil of $T_{A}^{\mathrm{max}}\sim 3.6$\,keV on a Ge nucleus. A quenching factor of $\sim 0.2$ then leads to a maximal ionisation signal of $E_{I}^{\rm max}\sim 750$\, eV.}
The coherent nature of this interaction channel enhances potential BSM contributions and renders it a strong probe for new physics.
To further underline its strong potential for future BSM studies, we show the expected CE$\nu$NS event rate dependence on the detector mass and threshold in figure~\ref{fig:signal_threshold_mass}.
While experimental exposure (mass $\cdot$ runtime) increases the event rate only linearly, improvements in the detector performance in terms of lower detection threshold and detection efficiency lead to a non-linear signal increase.
Experimental development in this direction will, therefore, be of great benefit to future investigations.
In actual experimental realisations, the expected signal might be further enhanced by scaling detector mass and measurement time.
Therefore, figure~\ref{fig:signal_threshold_mass} highlights the importance of future detector development for individual devices and quantifies the impact of further improvements in terms of the expected CE$\nu$NS signal.

%%%%%%%%%%%%%%%%%%%%%%%%%%%%%%%%%%%%%%%%%%%%%%%%%%%%%%%%%%%%%%%%%%%%%%%%%%%%%%%%%%%%%%%%%%%%%%%%%%%%

\begin{figure}[!t]
    \centering
    \includegraphics[width=0.48\textwidth]{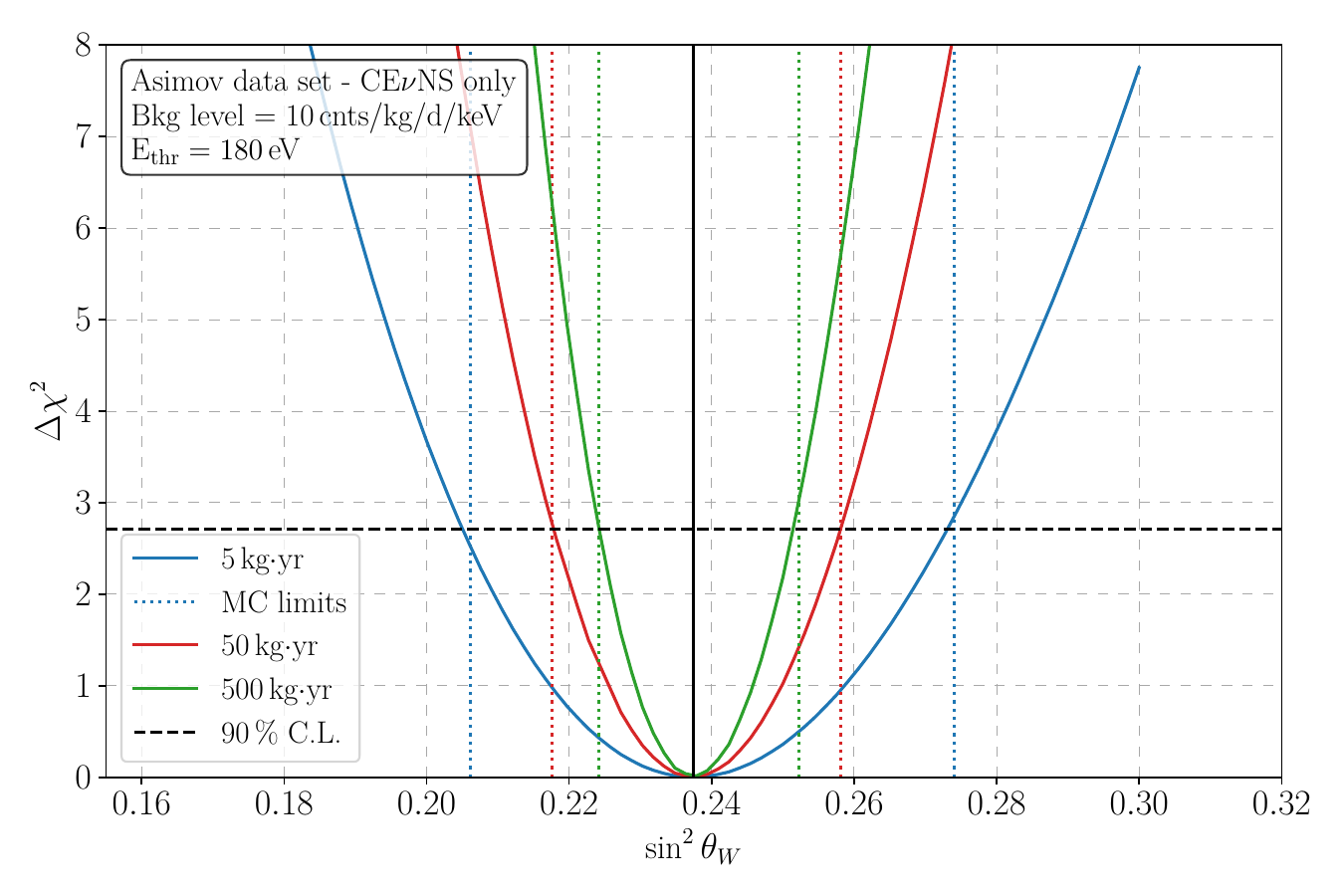} ~~
    \includegraphics[width=0.48\textwidth]{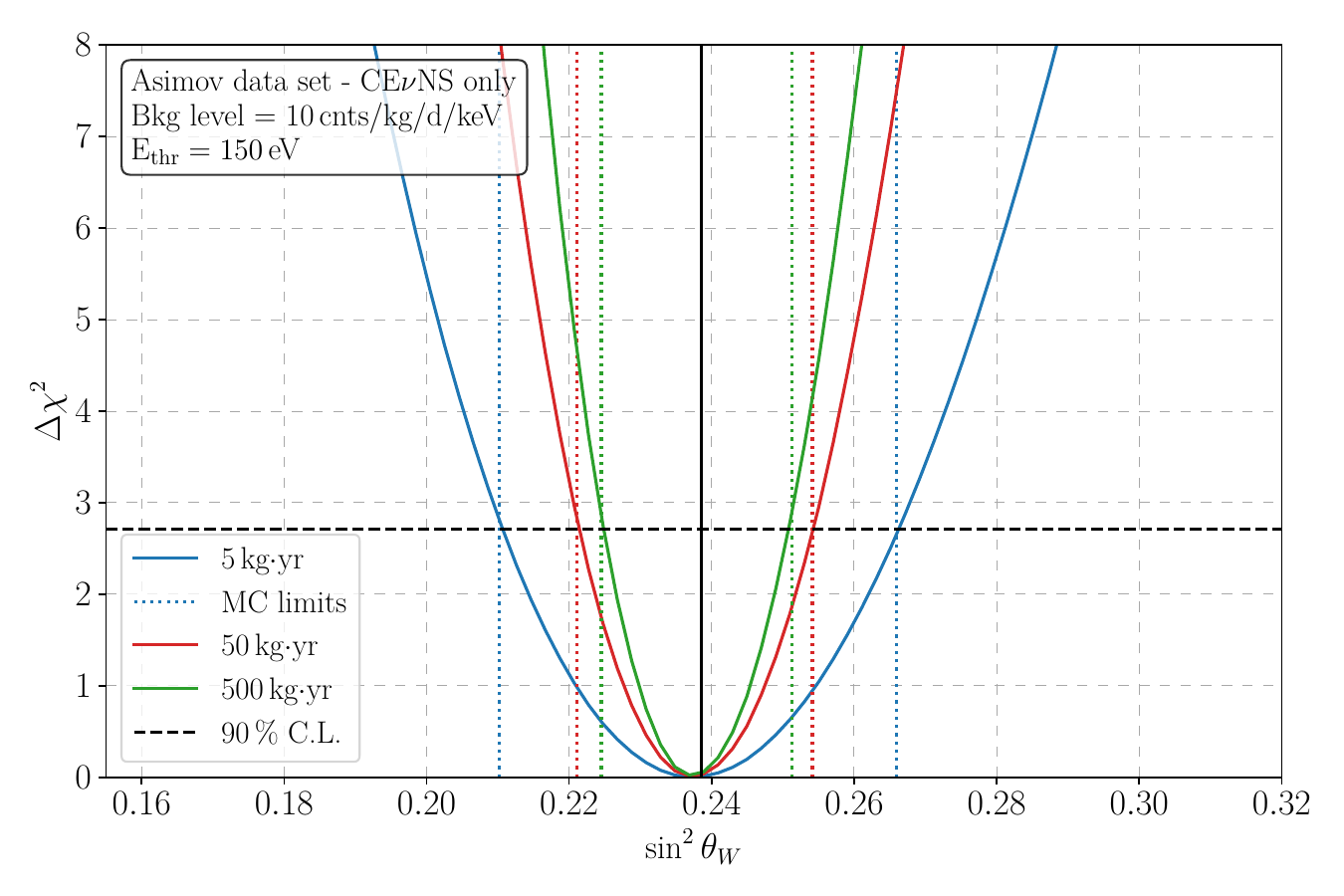} \\
    \includegraphics[width=0.48\textwidth]{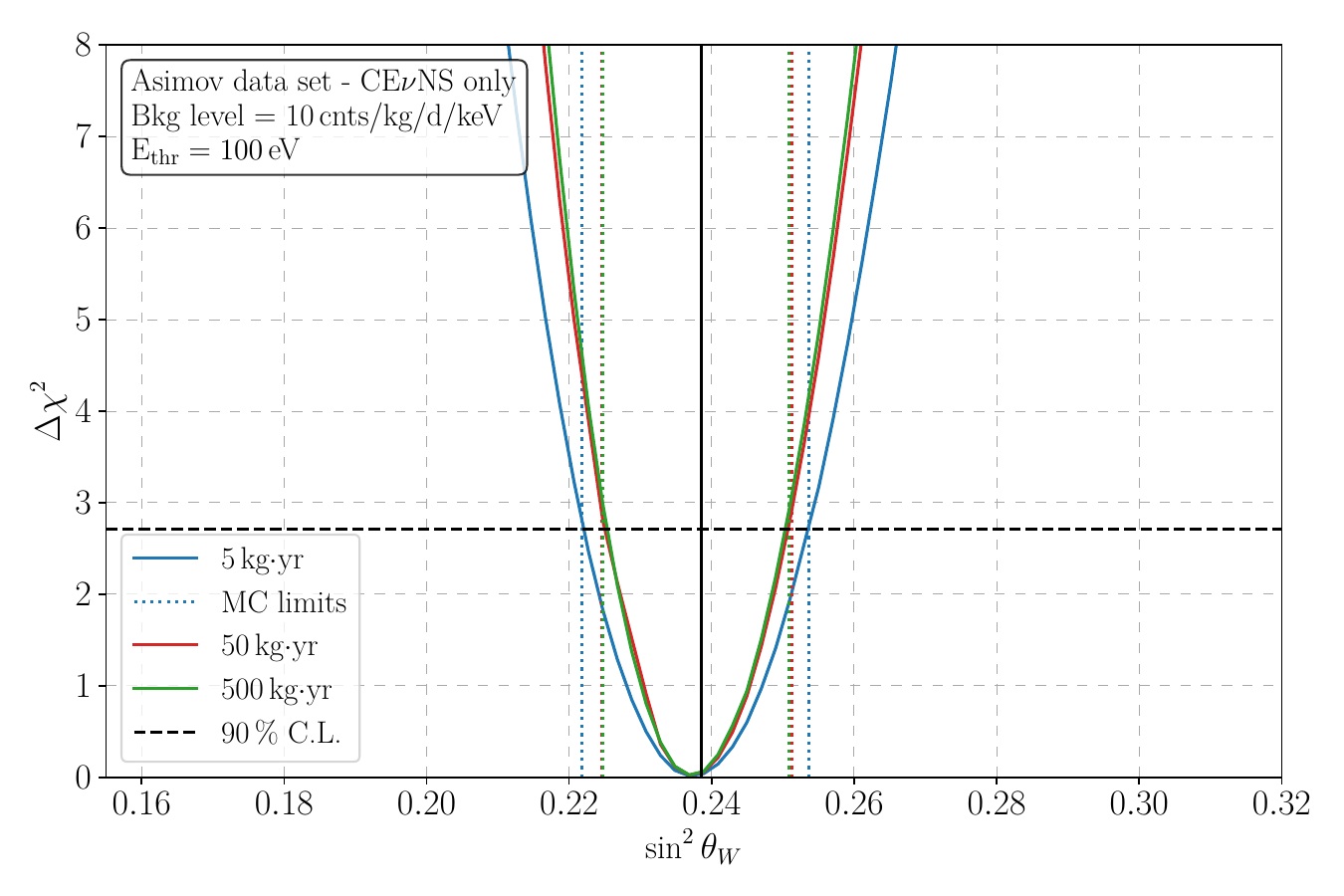}
    \caption{$\Delta \chi^{2}$ contours for the assumed experimental specifications (Asimov data sets). Background levels and detector thresholds are given in the boxes, while increasing exposures are shown in different colours. Assumptions on the background are similar to what has been achieved in the \textsc{Conus} experiment~\cite{CONUS:2020skt,CONUS:2022qbb}. The corresponding limits at 90\% C.L. are given by the intersection with the dashed line. Limits obtained from MC-sampled data are indicated by the dotted line.}
    \label{fig:cenns_weinberg_limits}
\end{figure}

%%%%%%%%%%%%%%%%%%%%%%%%%%%%%%%%%%%%%%%%%%%%%%%%%%%%%%%%%%%%%%%%%%%%%%%%%%%%%%%%%%%%%%%%%%%%%%%%%%%%

\begin{table}[]
    \centering
    \begin{tabular}{c|ccc}\hline
Exposure / $E_{\mathrm{thr}}$ & 180~eV & 150~eV & 100~eV \\ \hline
        5~kg$\cdot$yr   & (0.205, 0.273) & (0.211, 0.266)  & (0.222, 0.254) \\
            & (-14, 15)~\% & (-11, 12)~\%  & $\pm$7~\% \\
        50~kg$\cdot$yr  & (0.218, 0.258) & (0.221, 0.254)  & (0.225, 0.251) \\
            & (-8, 9)~\% & $\pm7$~\%  & (-5, 6)~\% \\
        500~kg$\cdot$yr & (0.224, 0.251) & (0.225, 0.251)  & (0.225, 0.25) \\
            & $\pm6$~\% & (-5, 6)~\%  & (-5, 6)~\% \\\hline
    \end{tabular}
    \caption{Error bands at 90\% C.L.\ and relative uncertainties of the weak mixing angle $\sin^{2}\theta_{W}$ for the different exposures and detection thresholds under study.}
    \label{tab:weinberg_contours}
\end{table}

%%%%%%%%%%%%%%%%%%%%%%%%%%%%%%%%%%%%%%%%%%%%%%%%%%%%%%%%%%%%%%%%%%%%%%%%%%%%%%%%%%%%%%%%%%%%%%%%%%%%

\subsection{Sensitivity to the weak mixing angle \texorpdfstring{$\sin^{2}\theta_{W}$}{}}

%%%%%%%%%%%%%%%%%%%%%%%%%%%%%%%%%%%%%%%%%%%%%%%%%%%%%%%%%%%%%%%%%%%%%%%%%%%%%%%%%%%%%%%%%%%%%%%%%%%%

\begin{figure}[!t]
\centering
    \includegraphics[width=0.48\textwidth]{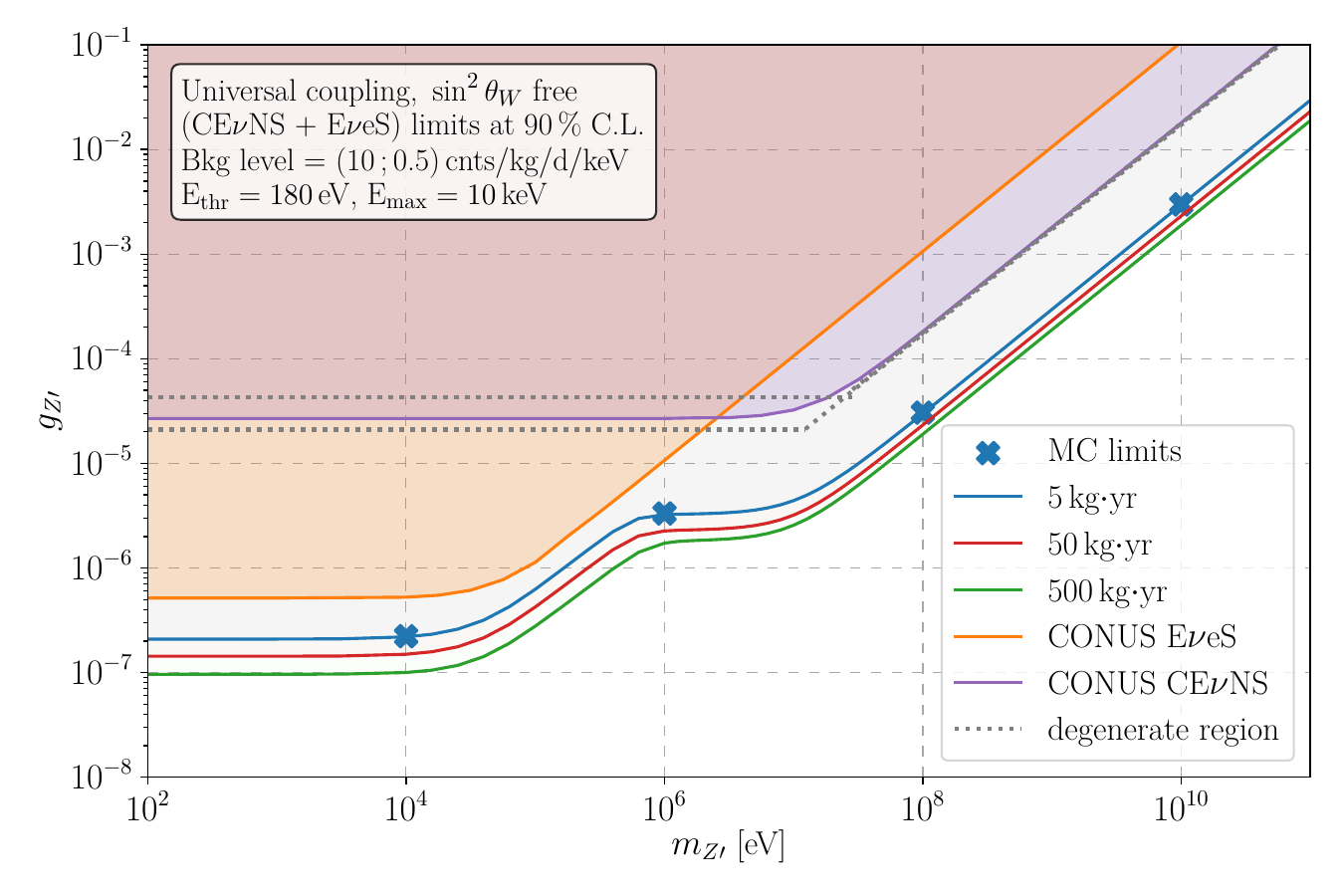} ~~
    \includegraphics[width=0.48\textwidth]{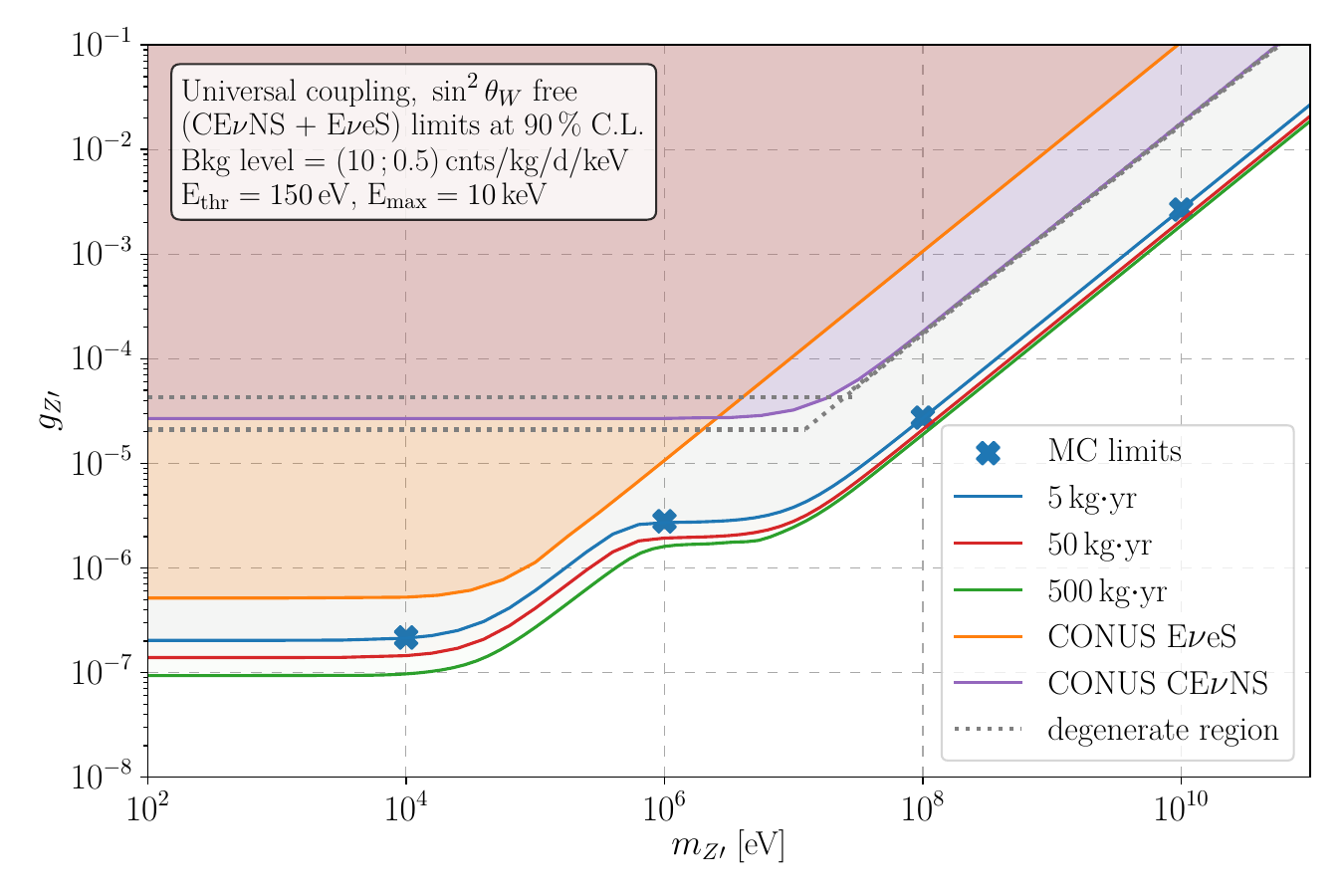}\\
    \includegraphics[width=0.48\textwidth]{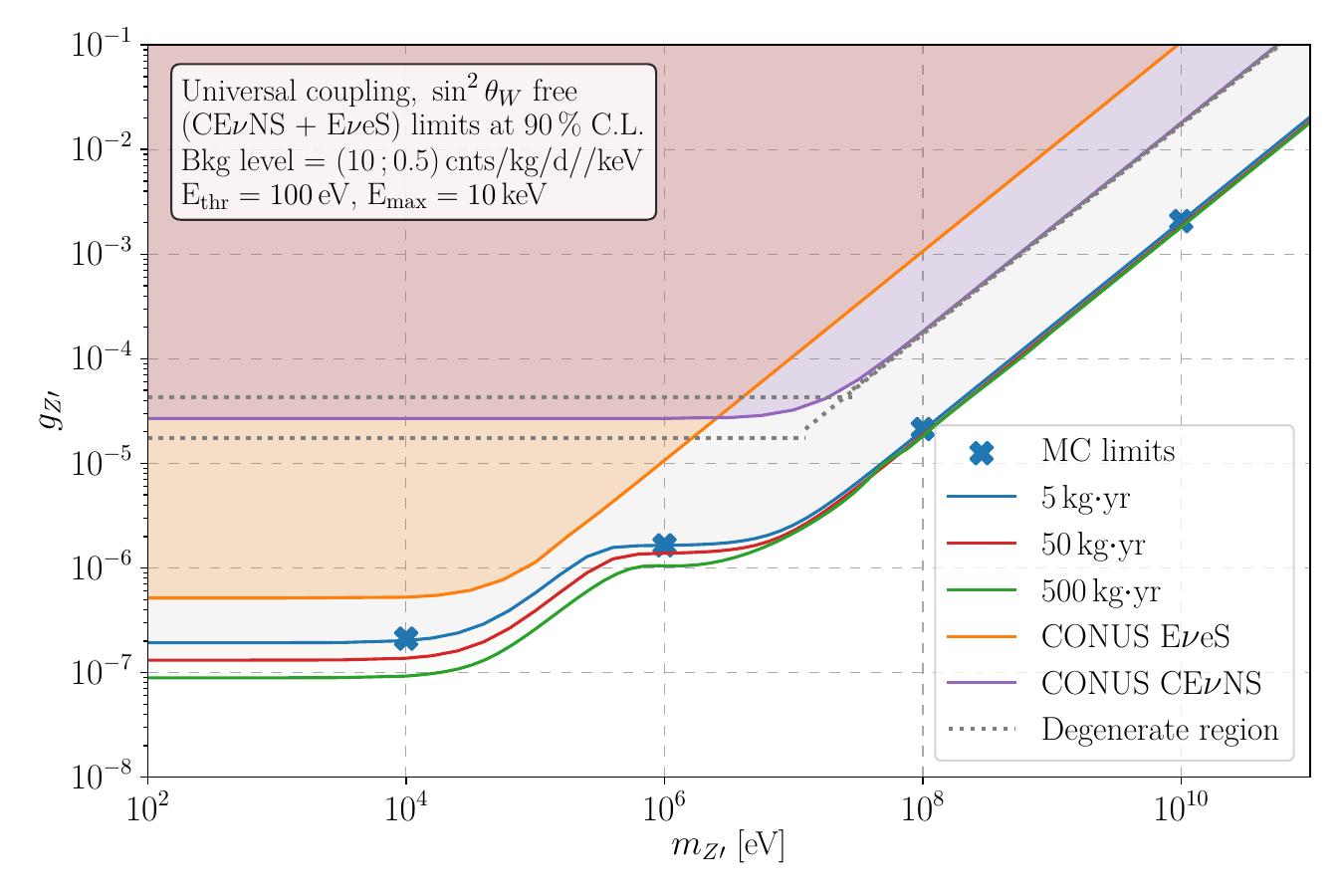}
    \caption{Expected exclusion limits at 90 \% C.L. on a universally coupled light vector mediator for different exposures and detector thresholds. Background levels and detector thresholds are given in the boxes, while the exposures under study are illustrated with different colouring. Exemplary limits from MC-sampled data are represented by crosses to validate our results. For comparison \textsc{Conus} limits from their first BSM investigation are indicated in orange (E$\nu$eS-only) and purple (CE$\nu$NS-only)~\cite{CONUS:2021dwh}. The degeneracy region for CE$\nu$NS, where the BSM model cannot be distinguished from the pure SM interaction, is marked with grey-dotted lines.}
    \label{fig:light_vector_limits_universal}
\end{figure}

%%%%%%%%%%%%%%%%%%%%%%%%%%%%%%%%%%%%%%%%%%%%%%%%%%%%%%%%%%%%%%%%%%%%%%%%%%%%%%%%%%%%%%%%%%%%%%%%%%%%

\begin{table}[]
    \resizebox{\textwidth}{!}{
    \centering
    \begin{tabular}{c|ccc}\hline
        Exposure / $E_{\mathrm{thr}}$ & 180~eV & 150~eV & 100~eV \\ \hline
        5~kg$\cdot$yr   & $2.1\cdot10^{-7}$ / $3.0\cdot10^{-4}$ & $2.0\cdot10^{-7}$ / $2.7\cdot10^{-4}$ & $1.9\cdot10^{-7}$ / $2.1\cdot10^{-4}$\\
        50~kg$\cdot$yr  & $1.4\cdot10^{-7}$ / $2.3\cdot10^{-4}$ & $1.4\cdot10^{-7}$ / $2.1\cdot10^{-4}$ & $1.3\cdot10^{-7}$ / $1.9\cdot10^{-4}$ \\
        500~kg$\cdot$yr & $9.7\cdot10^{-8}$ / $1.9\cdot10^{-4}$ & $9.4\cdot10^{-8}$ / $1.9\cdot10^{-4}$ & $9.0\cdot10^{-8}$ / $1.9\cdot10^{-4}$ \\ \hline
    \end{tabular}
    }
    \caption{Exemplary exclusion limits at 90\% C.L.\ for a universally coupled light vector boson. Values are given for regions where the experimental reach is dominated by E$\nu$eS ($m_{Z'}=1$\,keV) and CE$\nu$NS ($m_{Z'}=1$\,GeV), respectively.}
    \label{tab:light_vector_universal_limits}
\end{table}

%%%%%%%%%%%%%%%%%%%%%%%%%%%%%%%%%%%%%%%%%%%%%%%%%%%%%%%%%%%%%%%%%%%%%%%%%%%%%%%%%%%%%%%%%%%%%%%%%%%%

Since both CE$\nu$NS and E$\nu$eS cross sections exhibit a dependence on the weak mixing angle, these interaction channels can in principle be utilised for its investigation.
However, limits expected from E$\nu$eS alone are not competitive to existing bounds from electron scatterings.
Moreover, a combined analysis via both SM interaction channels is mainly driven by CE$\nu$NS.
Thus, we mainly focus on CE$\nu$NS and only present results obtained from energy regions below 1~keV. 
The outcome of our sensitivity estimation is shown in figure~\ref{fig:cenns_weinberg_limits} in terms of $\Delta\chi^{2}$ contours for the detection thresholds and experimental exposures under study.

%%%%%%%%%%%%%%%%%%%%%%%%%%%%%%%%%%%%%%%%%%%%%%%%%%%%%%%%%%%%%%%%%%%%%%%%%%%%%%%%%%%%%%%%%%%%%%%%%%%%

\begin{figure}[!t]
\centering
    \includegraphics[width=0.48\textwidth]{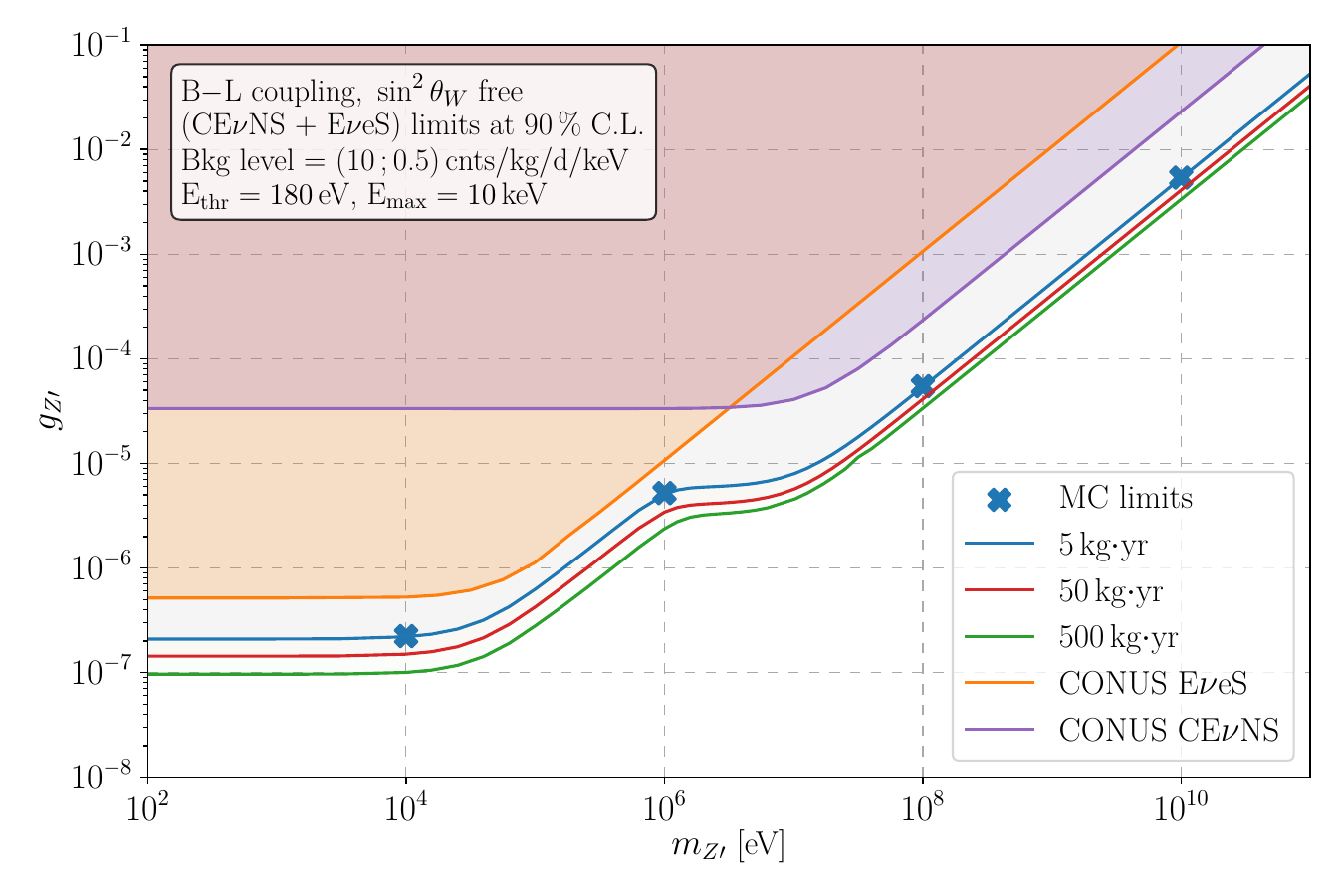} ~~
    \includegraphics[width=0.48\textwidth]{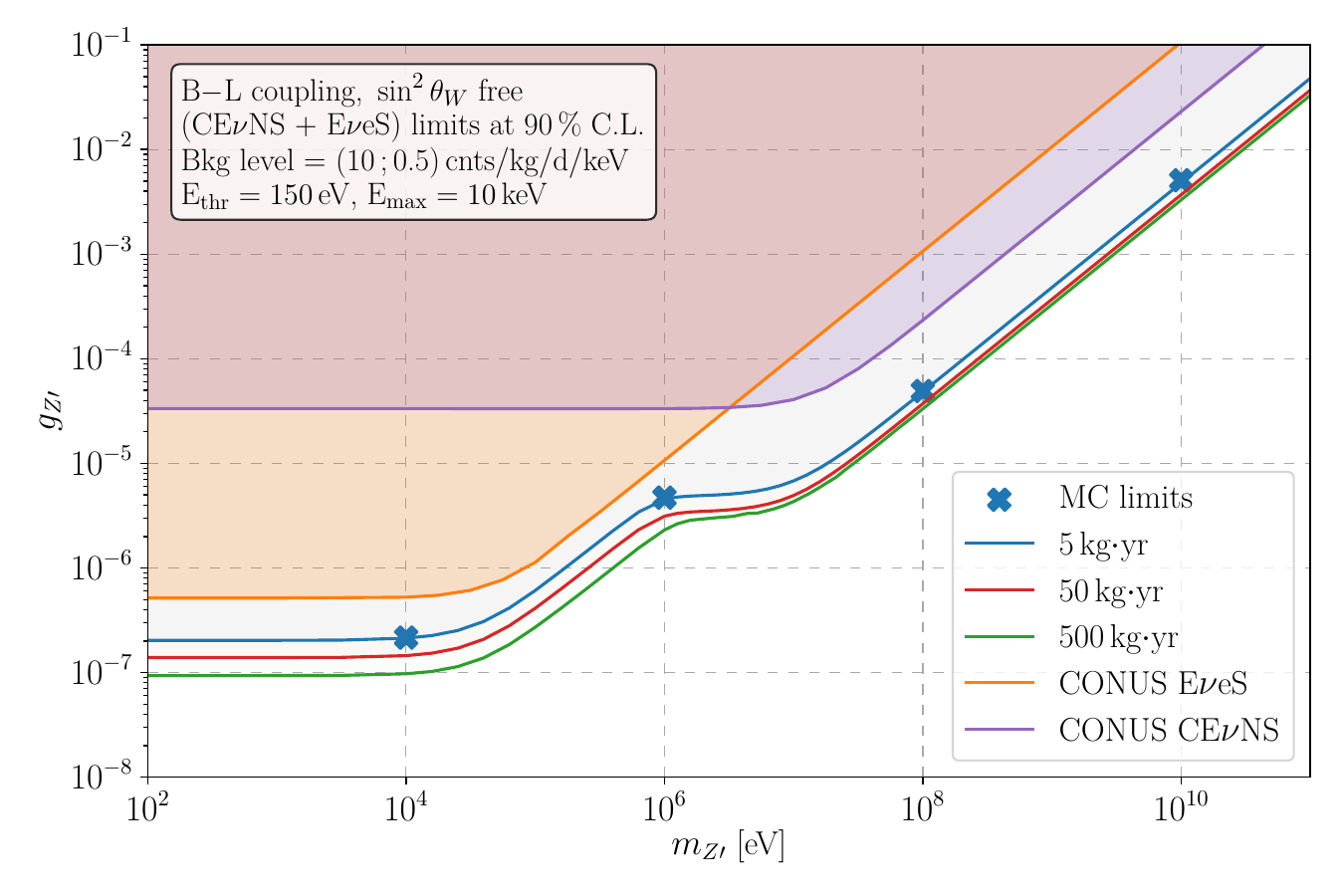} \\
    \includegraphics[width=0.48\textwidth]{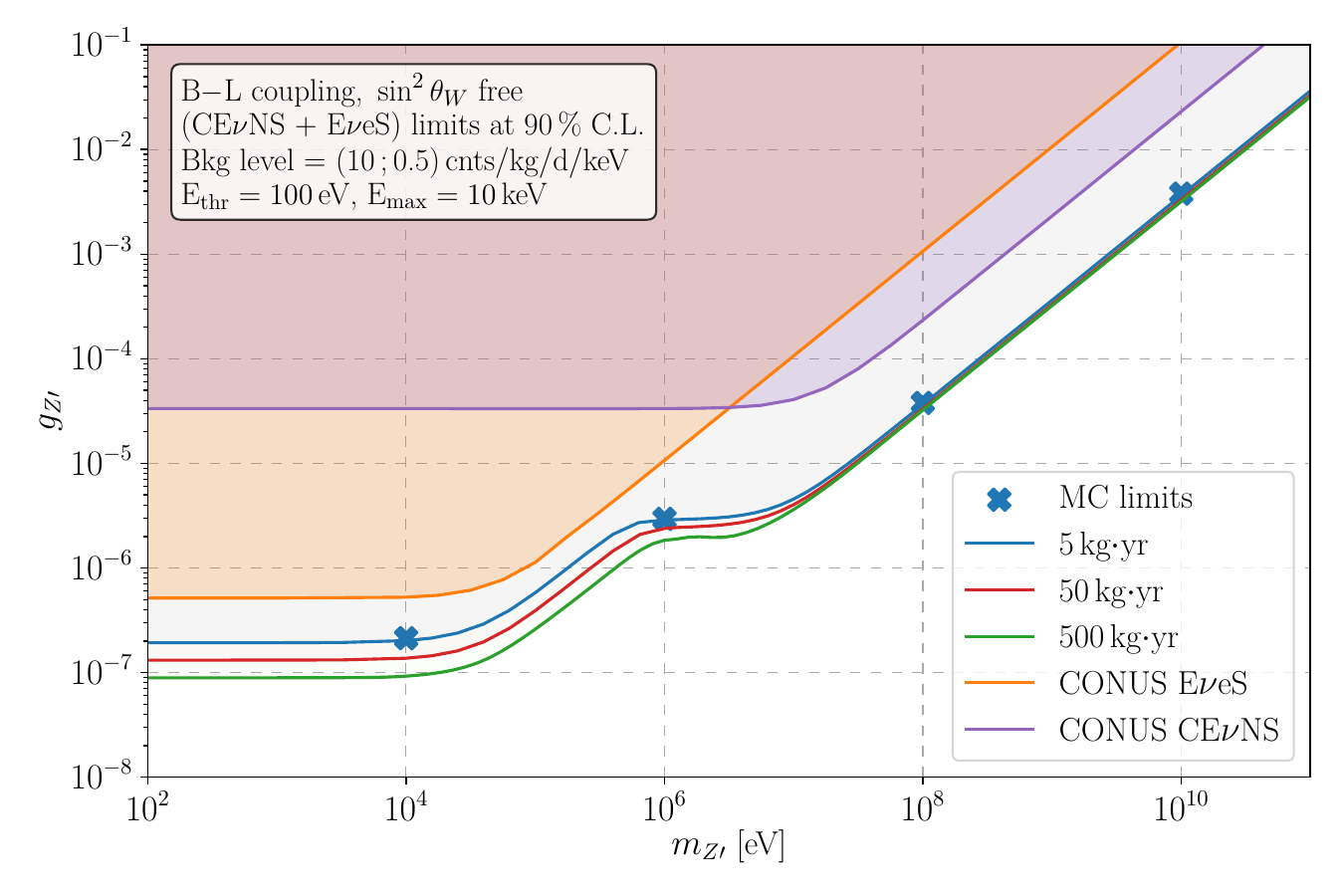}
    \caption{Expected exclusion limits at 90 \% C.L. on a light gauged B$-$L vector mediator for different exposures and detector thresholds. Background levels and detector thresholds are given in the boxes, while the exposures under study are illustrated with different colouring. Exemplary limits from MC-sampled data are represented by crosses to validate our results. For comparison \textsc{Conus} limits from their first BSM investigation are indicated in orange (E$\nu$eS-only) and purple (CE$\nu$NS-only)~\cite{CONUS:2021dwh}. In contrast to the previous case, the U(1)$_{\rm B-L}$ does not exhibit any degeneracy region.}
    \label{fig:light_vector_limits_b_l}
\end{figure}

%%%%%%%%%%%%%%%%%%%%%%%%%%%%%%%%%%%%%%%%%%%%%%%%%%%%%%%%%%%%%%%%%%%%%%%%%%%%%%%%%%%%%%%%%%%%%%%%%%%%

\begin{table}[]
\resizebox{\textwidth}{!}{
    \centering
    \begin{tabular}{c|ccc}\hline
        Exposure / $E_{\mathrm{thr}}$ & 180~eV & 150~eV & 100~eV \\ \hline
        5~kg$\cdot$yr   & $2.1\cdot10^{-7}$ / $5.3\cdot10^{-4}$ & $2.0\cdot10^{-7}$ / $4.8\cdot10^{-4}$ & $1.9\cdot10^{-7}$ / $3.6\cdot10^{-4}$\\
        50~kg$\cdot$yr  & $1.4\cdot10^{-7}$ / $4.1\cdot10^{-4}$ & $1.4\cdot10^{-7}$ / $3.7\cdot10^{-4}$ & $1.3\cdot10^{-7}$ / $3.3\cdot10^{-4}$ \\
        500~kg$\cdot$yr & $9.6\cdot10^{-8}$ / $3.4\cdot10^{-4}$ & $9.4\cdot10^{-8}$ / $3.3\cdot10^{-4}$ & $8.9\cdot10^{-8}$ / $3.2\cdot10^{-4}$ \\ \hline
    \end{tabular}
    }
    \caption{Exemplary exclusion limits at 90\% C.L.\ for a light U(1)$_{\rm B-L}$ vector boson. Values are given for regions where the experimental reach is dominated by E$\nu$eS ($m_{Z'}=1$\,keV) and CE$\nu$NS ($m_{Z'}=1$\,GeV), respectively.}
    \label{tab:light_vector_B_L_limits}
\end{table}

%%%%%%%%%%%%%%%%%%%%%%%%%%%%%%%%%%%%%%%%%%%%%%%%%%%%%%%%%%%%%%%%%%%%%%%%%%%%%%%%%%%%%%%%%%%%%%%%%%%%

The corresponding error bands at 90\% confidence level (C.L.) and relative uncertainties are given in table~\ref{tab:weinberg_contours}.
In general, results from Asimov data sets and MC-sampled mock data agree well.
As expected, the precision of the weak mixing angle increases with lower detection thresholds and increased experimental exposure.
While future experiments will determine the weak mixing angle at a level of $\lesssim \mathcal{O}(10)$~\%, underlying systematic uncertainties and the overall background level start to become more relevant for larger exposures and lower detection thresholds.
Hence, a detailed investigation of systematic uncertainties and their subsequent reduction will become crucial for future precision CE$\nu$NS experiments. 

%%%%%%%%%%%%%%%%%%%%%%%%%%%%%%%%%%%%%%%%%%%%%%%%%%%%%%%%%%%%%%%%%%%%%%%%%%%%%%%%%%%%%%%%%%%%%%%%%%%%

\subsection{Sensitivity to new light vector mediators}

Next, we present the expected sensitivity of future Ge-based reactor experiments to BSM physics in terms of potentially existing new light particles.
In this investigation, we simultaneously study BSM physics modifications to both CE$\nu$NS and E$\nu$eS as induced by the Lagrangian in eq.~\eqref{eq:lagrangian_light_vector}.
Thus, both channels are included in a combined likelihood function that is fit to Asimov/MC-sampled mock data.
Furthermore, the weak mixing angle is allowed to vary within the incorporated uncertainties, cf.\ section~\ref{sec:exp_spec}, while limits are determined for the vector coupling $g_{Z'}$ for a fixed boson mass $m_{Z'}$.
By repeating these fits, we can explore future sensitivities to the parameter space that is spanned by the model parameters by simultaneously allowing for potential changes in the weak mixing angle within current experimental uncertainties.
As a natural reference, we compare the expected sensitivities to CE$\nu$NS and E$\nu$eS limits obtained by the \textsc{Conus} experiments (after their conversion to our conventions)~\cite{CONUS:2021dwh}. 

%%%%%%%%%%%%%%%%%%%%%%%%%%%%%%%%%%%%%%%%%%%%%%%%%%%%%%%%%%%%%%%%%%%%%%%%%%%%%%%%%%%%%%%%%%%%%%%%%%%%

\paragraph{Universally coupled light vector boson}

The expected limits for the case of a universally coupled light vector boson are depicted in figure~\ref{fig:light_vector_limits_universal} with exemplary values given in table~\ref{tab:light_vector_universal_limits}.
The degenerate region of parameter space, i.e.\ the parameter configuration that is indistinguishable from the SM, is indicated by the grey bands.
Due to a lighter target mass, the obtained bounds in the keV-mass region are dominated by the electron channel, whereas CE$\nu$NS sets exclusion limits for the higher (GeV-)mass region.
Comparing to the respective \textsc{Conus} limits, i.e.,\  $5.2\cdot10^{-7}$ for E$\nu$eS-only ($m_{Z'}=1$\,keV), and $1.8\cdot10^{-3}$ for CE$\nu$NS-only ($m_{Z'}=1$\,GeV), future experimental efforts promise an improvement of roughly one order of magnitude.

%%%%%%%%%%%%%%%%%%%%%%%%%%%%%%%%%%%%%%%%%%%%%%%%%%%%%%%%%%%%%%%%%%%%%%%%%%%%%%%%%%%%%%%%%%%%%%%%%%%%

\paragraph{Light gauged B$-$L vector boson}

Our results for the case of the U(1)$_{\rm B-L}$ model are shown in figure~\ref{fig:light_vector_limits_b_l} with benchmark values given in table~\ref{tab:light_vector_B_L_limits}.
Since the corresponding cross-section in eq.~\eqref{eq:cross_section_vector_b_l} is well defined, the model's parameter space can be directly accessed.
The different charge assignment compared to the previous model leads to a modification of the cross section by a factor of $\sim3$ and, thus, slightly weaker bounds.
Besides that, the expected sensitivities are similar to the previous case.
The corresponding benchmark limits from \textsc{Conus} are $5.2\cdot10^{-7}$ for E$\nu$eS-only ($m_{Z'}=1$\,keV) and $2.3\cdot10^{-3}$ for CE$\nu$NS-only ($m_{Z'}=1$\,GeV), respectively.

Expected limits on the parameter space and corresponding benchmark points for both light vector models in the case of an extended high energy region, i.e.\ up to 100\,keV, are given in appendix~\ref{app:vector_limits_extended_energy}.

Finally, we emphasise the complementarity between experiments using reactor antineutrinos and those using (anti-)neutrinos from pion-decays-at-rest.
The huge antineutrino flux of nuclear reactors allows for strong restriction of the models' parameter spaces, especially via E$\nu$eS. The lower momentum transfer of reactor antineutrino also renders CE$\nu$NS more sensitive to lower mediator masses.  
On the other hand, higher (anti-)neutrino energies, as provided by $\pi$DAR sources, allow the probe of heavier particles and give them a higher sensitivity for the high mass region of the model studied, cf.\ right plot in figure~\ref{fig:money_plot}.
Thus, by combining data from both sources, the parameter space of new light vector particles can be efficiently probed with CE$\nu$NS detectors.

%%%%%%%%%%%%%%%%%%%%%%%%%%%%%%%%%%%%%%%%%%%%%%%%%%%%%%%%%%%%%%%%%%%%%%%%%%%%%%%%%%%%%%%%%%%%%%%%%%%%

\section{Conclusions}\label{sec:conclusions}

In this paper, the potential of future germanium-based reactor experiments was investigated, using the weak mixing angle $\sin^{2}\theta_{W}$ and new light vector bosons as examples.
In particular, the expected large signal of CE$\nu$NS renders such experiments promising tools for future SM and BSM investigations since it allows for ``car-size'' neutrino detectors close to the strongest artificial neutrino sources on Earth.
Although multiple confirmations of successful CE$\nu$NS detection are still lacking at a reactor site, this channel is currently probed with the full repertoire of modern detection technologies.
This study focused on the potential of germanium semiconductor detectors as these are currently leading the field at reactor sites.
The enormous antineutrino flux further allows obtaining competitive bounds using the electrons of the detector as target material via E$\nu$eS.

We investigated the expected sensitivity of potential set-ups for selected detection thresholds and experimental exposure, cf.~table~\ref{tab:experimental_configurations}, using the SM interactions channels CE$\nu$ES and E$\nu$eS. 
For these selected configurations, realistic detector efficiency and resolution are considered as well as systematic uncertainties related to reactor antineutrinos and signal quenching are taken into account.
The signal expectations of CE$\nu$NS and E$\nu$eS have been estimated and the impact of key experimental parameters like detector mass and detection threshold are quantified using the expected event rate of CE$\nu$NS as an example, cf.~figure~\ref{fig:signal_threshold_mass}. In doing so, the potential of future SM and BSM investigations has been highlighted.

\begin{figure}[t]
\centering
    \includegraphics[width=0.48\textwidth]{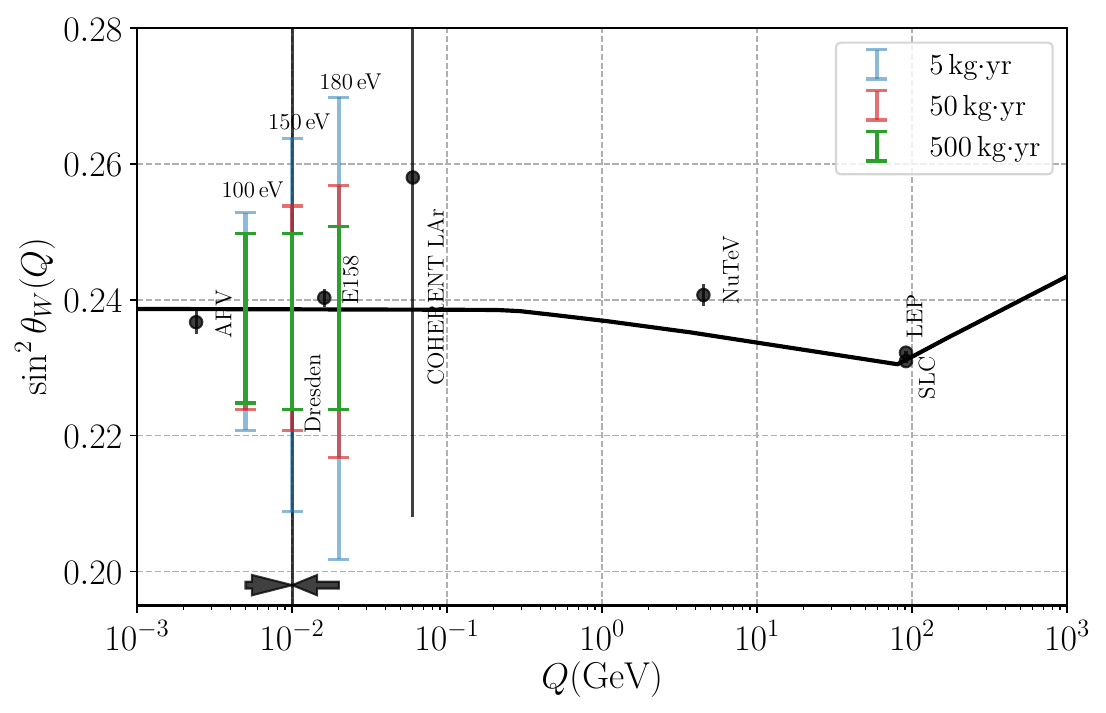}~~
    \includegraphics[width=0.48\textwidth]{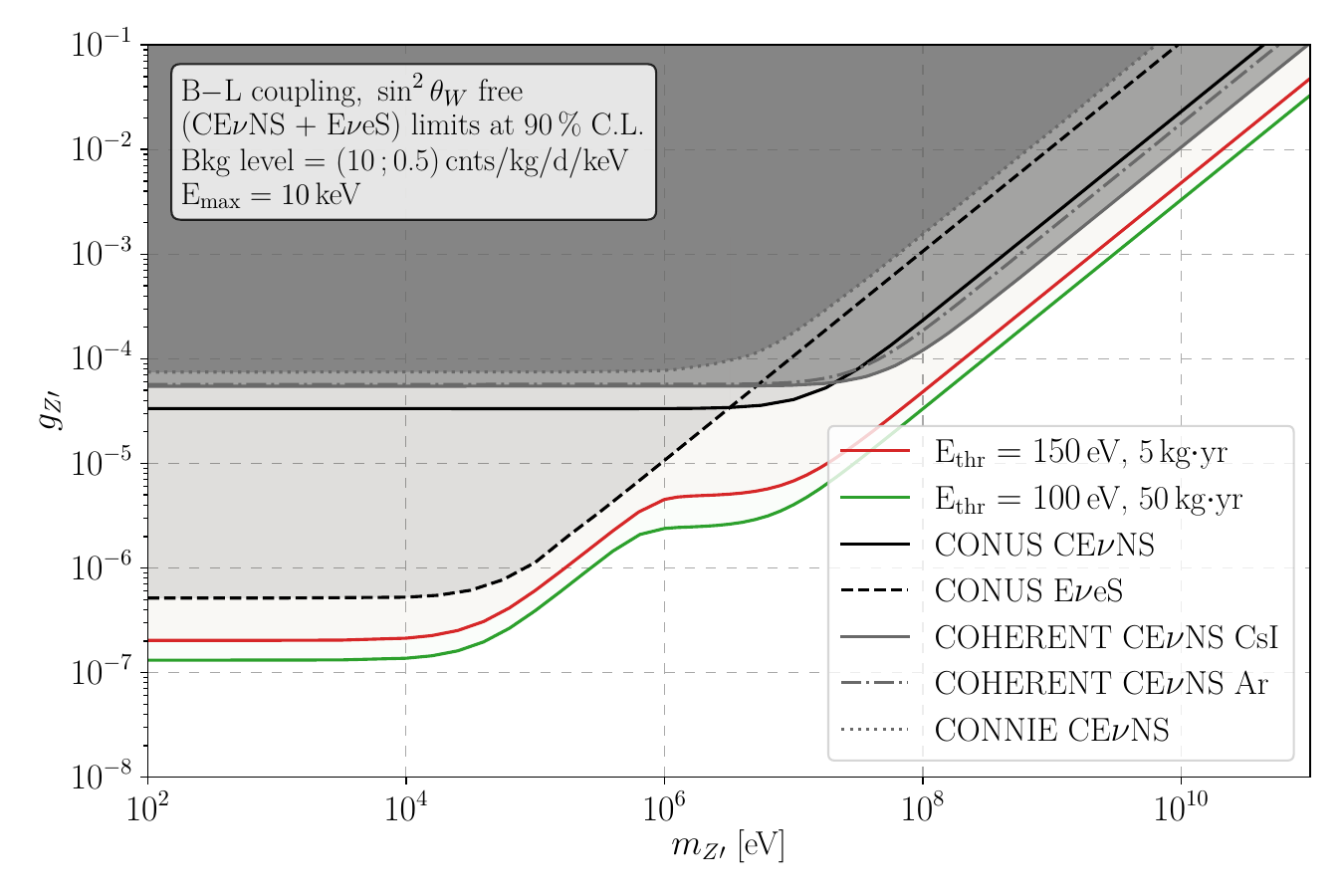}
    \caption{Summary of the investigations in this work. Left: Expected precision to the weak mixing angle $\sin^{2}\theta_{W}$ in comparison to existing measurements at different energy scales. CE$\nu$NS provides measurements via neutrinos in the low-momentum range of $10^{-2}$ - $10^{-1}$\,GeV, with future experiments capable of entering the precision regime (see text for more details). Right: Expected sensitivity to the parameter space of a gauged B$-$L vector boson in comparison to existing limits from CE$\nu$NS experiments at a $\pi$DAR source (\textsc{Coherent}) and a nuclear reactors (\textsc{Connie}, \textsc{Conus}).}
    \label{fig:money_plot}
\end{figure}

Further, we obtained that the relative uncertainty of the weak mixing angle is expected to improve to $\lesssim 10$~\%, cf.~figure~\ref{fig:cenns_weinberg_limits} and table~\ref{tab:weinberg_contours}. Our results and some of the current measurements are summarized in the left panel of figure~\ref{fig:money_plot}. The most precise measurements are determined at the Z-pole (LEP, SLC)~\cite{ALEPH:2005ab}. On the other hand, the lowest energy measurements are obtained by measurements of atomic parity violation (APV) with cesium~\cite{Wood:1997zq,Guena:2004sq}. Complementary measurements have also been performed in polarized M\o ller scatterings with electrons (E158)~\cite{MOLLER:2014iki}.
A precise direct determination of the weak mixing angle in neutrino scattering (at $Q\sim 4.5$\,GeV) comes from the NuTeV collaboration~\cite{NuTeV:2001whx}. More recently, measurements from CE$\nu$NS in the low-momentum range of $10^{-2}$ - $10^{-1}$\,GeV has been used to constrain the weak mixing angle~\cite{Cadeddu:2021ijh}.
We find that experiments like \textsc{Conus}+~\cite{Ackermann:2024kxo} are expected to have a better sensitivity to the weak mixing angle, at energies of around 10~MeV, than the current measurements by the \textsc{Coherent} and the Dresden experiments and, thus, representing an important step towards precision measurements of the weak mixing angle with neutrinos.

For the light vector models, bounds on the coupling strength $g_{Z'}$ strengthen by roughly one order of magnitude, depending on the explicit experimental configuration, cf.~figure~\ref{fig:light_vector_limits_universal} and table~\ref{tab:light_vector_universal_limits} for the universally coupled model as well as figure~\ref{fig:light_vector_limits_b_l} and table~\ref{tab:light_vector_B_L_limits} for the U(1)$_{\rm B-L}$ model. As a demonstration, we show the sensitivity of a future germanium experiment, such as \textsc{Conus}+, to the B$-$L model in the right panel of figure~\ref{fig:money_plot} for two different configurations of the experimental parameters that seem realisable in the near future. For comparison, the corresponding CE$\nu$NS bounds from the \textsc{Coherent} and \textsc{Connie} collaborations are also shown.\footnote{The limits on a universally coupled light vector mediator published by the \textsc{Connie} collaboration~\cite{CONNIE:2024pwt} have been rescaled to match the correct charge assignment of the B$-$L model, cf.\,section~\ref{sec:channels}.} The Skipper-CCD technology used by the \textsc{Connie} collaboration provides competitive bounds in the low mediator mass region since its exceptionally low detection threshold of $\mathcal{O}(10)$\,eV compensates for their low detector mass ($<1$\,g/sensor). On the other side, germanium semiconductor experiments have the capability to offset the elevated detection thresholds through increasing the overall detector mass with more germanium crystals ($\sim1$\,kg/diode). Nevertheless, both technologies represent interesting paths towards future CE$\nu$NS measurements close to a nuclear reactor.

%%%%%%%%%%%%%%%%%%%%%%%%%%%%%%%%%%%%%%%%%%%%%%%%%%%%%%%%%%%%%%%%%%%%%%%%%%%%%%%%%%%%%%%%%%%%%%%%%%%%

%%%%%%%%%%%%%%%%%%%%%%%%%%%%%%%%%%%%%%%%%%%%%%%%%%%%%%%%%%%%%%%%%%%%%%%%%%%%%%%%%%%%%%%%%%%%%%%%%%%%

We found that limits on the E$\nu$eS ($m_{Z'}=1$\,keV) region are stronger driven by increasing exposure due to the signal's flatness.
On the contrary, limits in the CE$\nu$NS ($m_{Z'}=1$\,GeV) region benefit stronger from an improved detection threshold.
In general, the combination of both SM channels in one analysis has proven to be a very efficient way of accessing the available parameter space.
Furthermore, incorporating data from $\pi$DAR CE$\nu$NS experiments will improve bounds for higher mediators masses.
Thus, exploiting this reactor-$\pi$DAR complementarity in future investigations is a sufficient way of constraining the full parameter space.

This study further underlines the large potential of high-purity point-contact germanium detectors, also in critical, non-laboratory environments.
Ongoing improvements in detector resolution, detection threshold and detection efficiency in the sub-keV region will increase the expected CE$\nu$NS events rates and consequently the reach to any potentially existing BSM physics related to it.
In addition, the scaling capability of this detection technology further allows to linearly increase the signal rates.
In that context, background and environmental stability are key to guarantee this initial potential.

The initially mentioned experimental attempts to measure CE$\nu$NS close to a nuclear reactor use different detection technologies with their own characteristics in terms of expected backgrounds, efficiencies and threshold energies. Intrinsic material backgrounds affect the final sensitivity as well as individual shielding capabilities. Devices that achieve very low detection thresholds might tolerate higher background levels due to an enhanced CE$\nu$NS signal. Consequently, experiments with higher detection thresholds are more subject to critical backgrounds and efficiency requirements. Moreover, data acquisition and analysis of CE$\nu$NS and E$\nu$eS events usually depends on the applied technology and comes usually with different detection efficiencies and sensitivities for both channels.

With more small-scale technologies showing their capability of measuring CE$\nu$NS both close to reactors and at $\pi$DAR sources, neutrino physics enters a high-statistics phase when their full detector and scaling potential is exploited.
Consequently, SM and BSM reaches of CE$\nu$NS experiments will start to be dominated by the systematic (experimental) uncertainties which call for a stronger consideration in future sensitivity studies. 
Beyond that, the current generation of dark matter direct detection experiments is beginning to probe the neutrino fog, e.g.\ $\sim2$\,cnts/(0.6\,t$\cdot$yr) from CE$\nu$NS of solar $^{8}$B neutrinos have already been expected in the \textsc{Xenon1T} experiment~\cite{XENON:2020gfr}. Recently reported results~\cite{XENON:2024solar} show first signs of this signal in \textsc{Xenon}n\textsc{T}, which demonstrates that the sun will become available as additional neutrino source in future CE$\nu$NS studies.
In summary, the prospects for future CE$\nu$NS investigations are promising and there will be a huge playground for future phenomenological studies, especially if experimenters and theorists work closely together.

%%%%%%%%%%%%%%%%%%%%%%%%%%%%%%%%%%%%%%%%%%%%%%%%%%%%%%%%%%%%%%%%%%%%%%%%%%%%%%%%%%%%%%%%%%%%%%%%%%%%

\acknowledgments

The authors would like to thank the \textsc{Conus} collaboration for important insights into current developments of germanium semiconductor detectors. Special thanks are given to Aurélie Bonhomme, Janina Hakenmüller and Edgar Sanchez Garcia for useful discussions.

%%%%%%%%%%%%%%%%%%%%%%%%%%%%%%%%%%%%%%%%%%%%%%%%%%%%%%%%%%%%%%%%%%%%%%%%%%%%%%%%%%%%%%%%%%%%%%%%%%%%

\appendix 

\section{Limits on new light vector bosons with an extended high energy range}\label{app:vector_limits_extended_energy}

In the following, we summarise results on the parameter space of the investigated light vector models with an increased energy region.
For all plots and tables below, the upper end of energy region was increased from 10\,keV to 100\,keV. 
As a result, the effect of E$\nu$eS on the expected sensitivity increases, and the limit contours and the given reference points change slightly.
Furthermore, individual model characteristics become more apparent: A gauged B$-$L model only contributes additional events in both CE$\nu$NS and E$\nu$sS, while there is some interplay between both channels in the case of a universally coupled light mediators as CE$\nu$NS events are also reduced in a certain region of parameter space.

%%%%%%%%%%%%%%%%%%%%%%%%%%%%%%%%%%%%%%%%%%%%%%%%%%%%%%%%%%%%%%%%%%%%%%%%%%%%%%%%%%%%%%%%%%%%%%%%%%%%

\subsection{Universally coupled light vector boson}

%%%%%%%%%%%%%%%%%%%%%%%%%%%%%%%%%%%%%%%%%%%%%%%%%%%%%%%%%%%%%%%%%%%%%%%%%%%%%%%%%%%%%%%%%%%%%%%%%%%%

\begin{figure}[H]
\centering
    \includegraphics[width=0.48\textwidth]{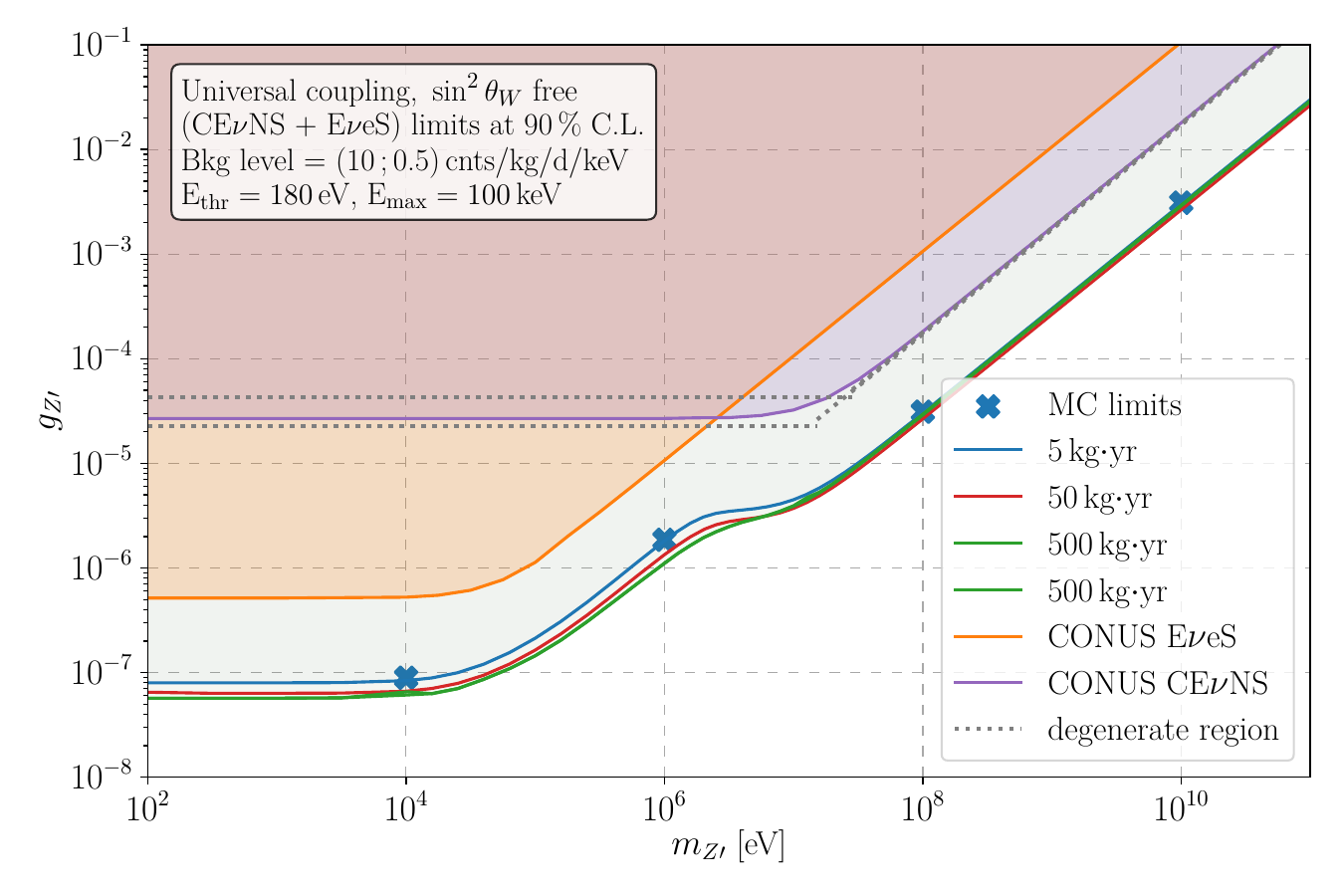} ~~
    \includegraphics[width=0.48\textwidth]{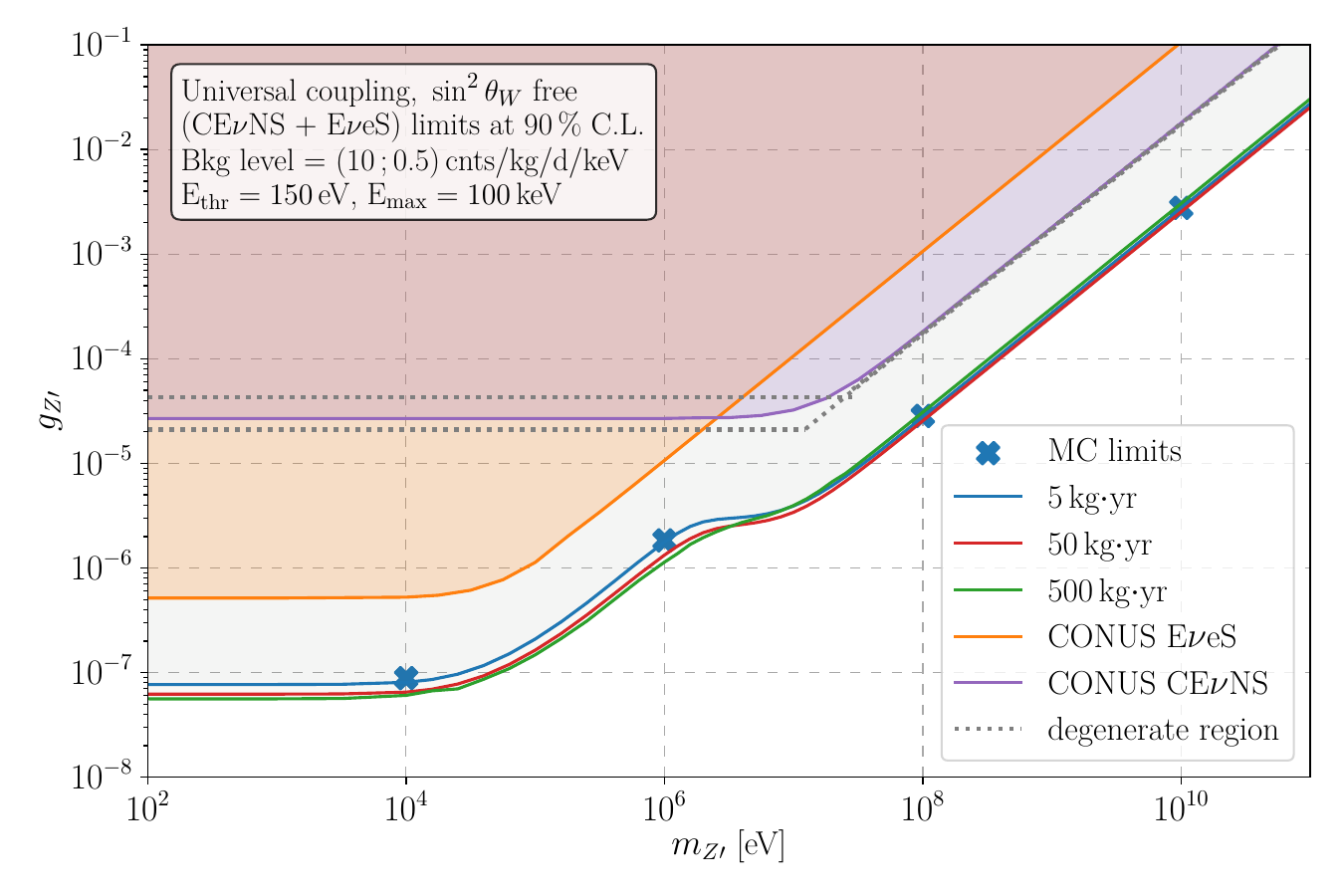} \\
    \includegraphics[width=0.48\textwidth]{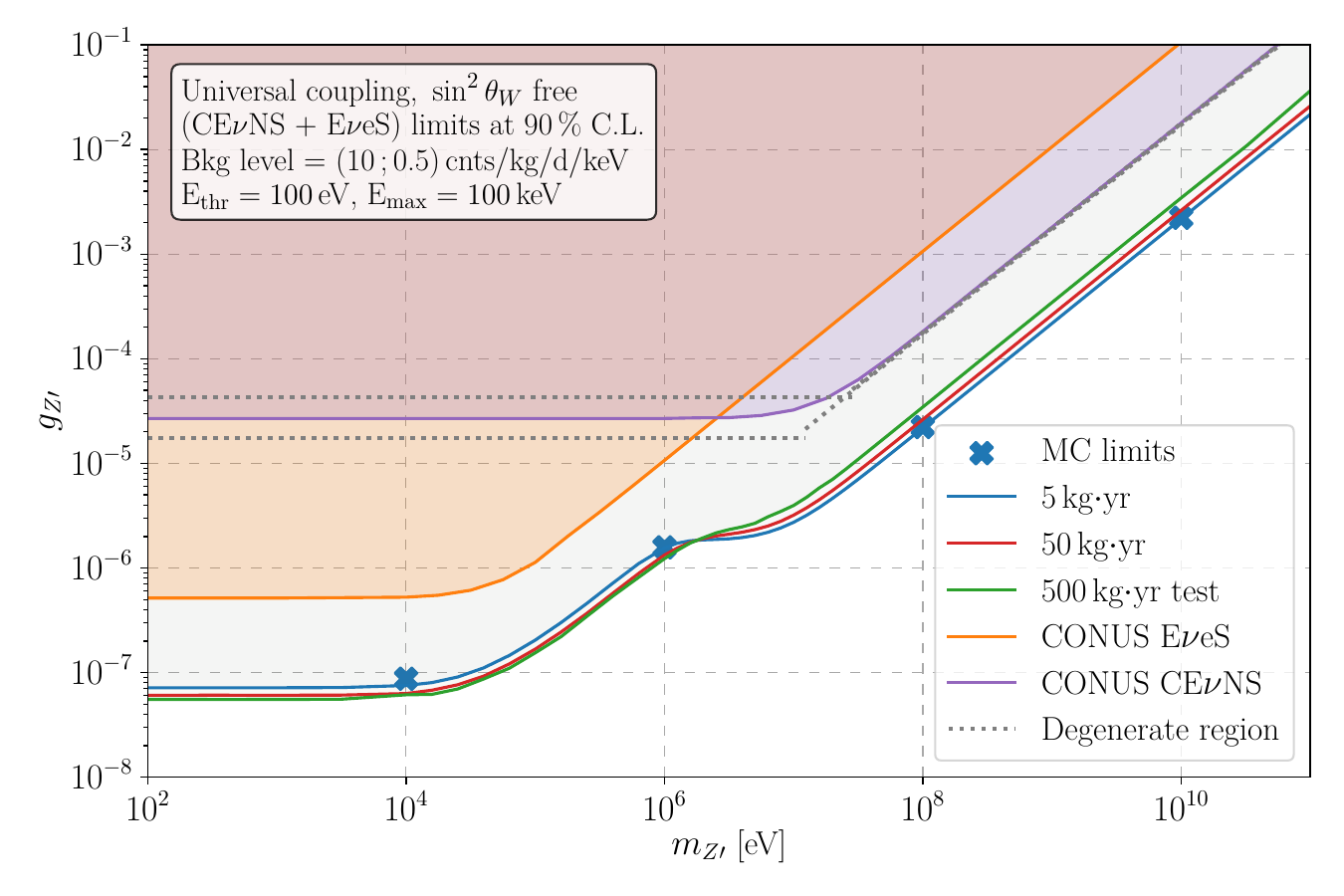}
    \caption{Expected exclusion limits at 90 \% C.L. on a universally coupled light vector mediator for an extended energy region ($E_{I}\leq100$\,keV) and different exposures and detector thresholds. Background levels and detector thresholds are given in the boxes, while the exposures under study are illustrated with different colouring. Exemplary limits from MC-sampled data are represented by crosses to validate our results. For comparison \textsc{Conus} limits from their first BSM investigation are indicated in orange (E$\nu$eS-only) and purple (CE$\nu$NS-only)~\cite{CONUS:2021dwh}. The degeneracy region for CE$\nu$NS, where the BSM model cannot be distinguished from the pure SM interaction, is marked with grey-dotted lines.}
    \label{fig:light_vector_limits_universal_long}
\end{figure}

%%%%%%%%%%%%%%%%%%%%%%%%%%%%%%%%%%%%%%%%%%%%%%%%%%%%%%%%%%%%%%%%%%%%%%%%%%%%%%%%%%%%%%%%%%%%%%%%%%%%

\begin{table}[H]
\resizebox{\textwidth}{!}{

    \centering
    \begin{tabular}{c|ccc}\hline
        Exposure / $E_{\mathrm{thr}}$ & 180~eV & 150~eV & 100~eV \\ \hline
        5~kg$\cdot$yr   & $8.0\cdot10^{-8}$ / $3.0\cdot10^{-4}$ & $7.7\cdot10^{-8}$ / $2.8\cdot10^{-4}$ & $7.2\cdot10^{-8}$ / $2.2\cdot10^{-4}$\\
        50~kg$\cdot$yr  & $6.3\cdot10^{-8}$ / $2.7\cdot10^{-4}$ & $6.2\cdot10^{-8}$ / $2.5\cdot10^{-4}$ & $6.1\cdot10^{-8}$ / $2.6\cdot10^{-4}$ \\
        500~kg$\cdot$yr & $5.7\cdot10^{-8}$ / $2.9\cdot10^{-4}$ & $5.6\cdot10^{-8}$ / $3.1\cdot10^{-4}$ & $5.5\cdot10^{-8}$ / $3.5\cdot10^{-4}$ \\ \hline
    \end{tabular}
    }
    \caption{Exemplary exclusion limits at 90\% C.L.\ for a universally coupled light vector boson for an extended energy range ($E_{I}\leq100$\,keV). Values are given for regions where the experimental reach is dominated by E$\nu$eS ($m_{Z'}=1$\,keV) and CE$\nu$NS ($m_{Z'}=1$\,GeV), respectively.}
    \label{tab:light_vector_universal_limits_long}
\end{table}

%%%%%%%%%%%%%%%%%%%%%%%%%%%%%%%%%%%%%%%%%%%%%%%%%%%%%%%%%%%%%%%%%%%%%%%%%%%%%%%%%%%%%%%%%%%%%%%%%%%%

\subsection{Light gauged B$-$L vector boson}

%%%%%%%%%%%%%%%%%%%%%%%%%%%%%%%%%%%%%%%%%%%%%%%%%%%%%%%%%%%%%%%%%%%%%%%%%%%%%%%%%%%%%%%%%%%%%%%%%%%%

\begin{figure}[H]
\centering
    \includegraphics[width=0.48\textwidth]{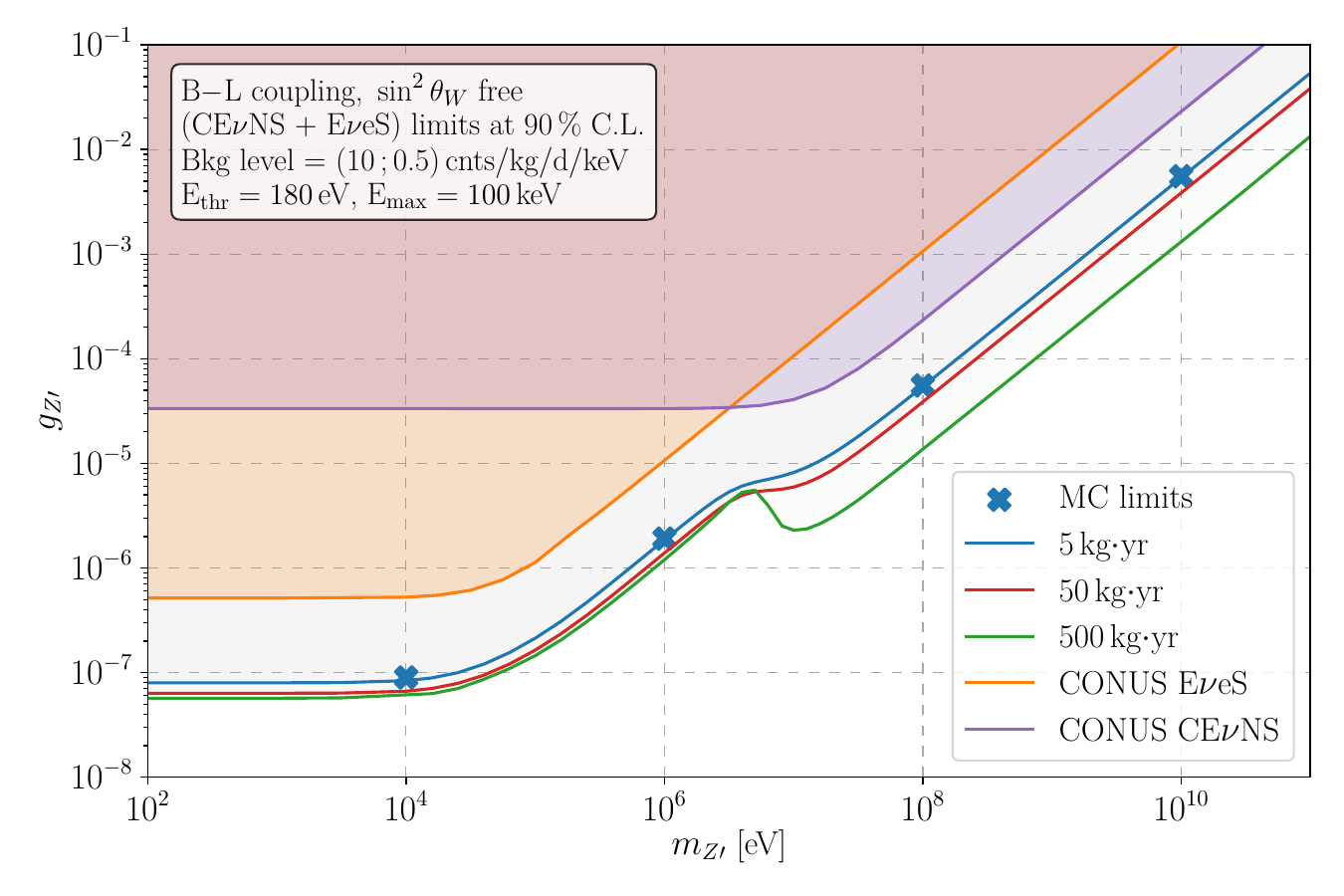} ~~
    \includegraphics[width=0.48\textwidth]{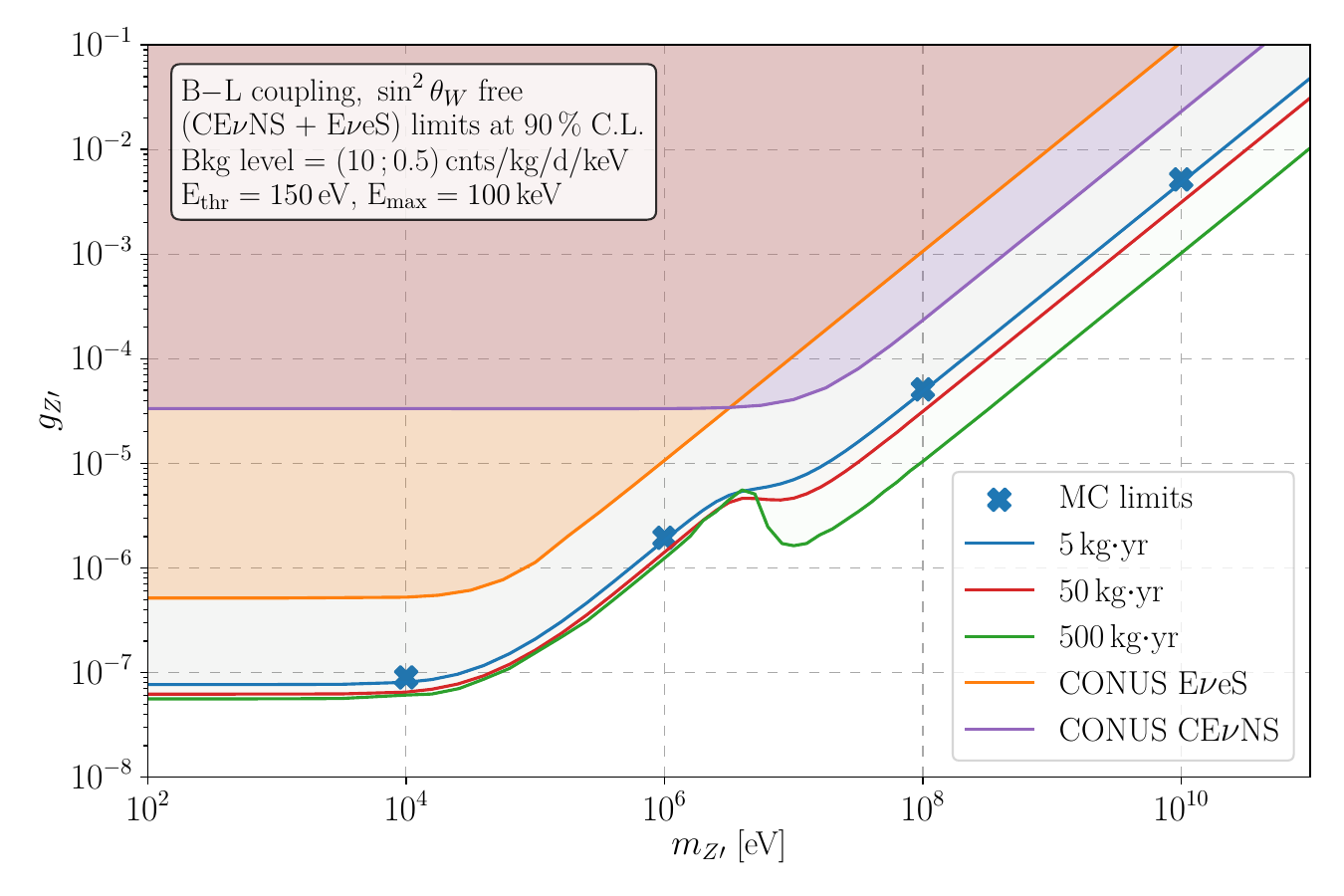} \\
    \includegraphics[width=0.48\textwidth]{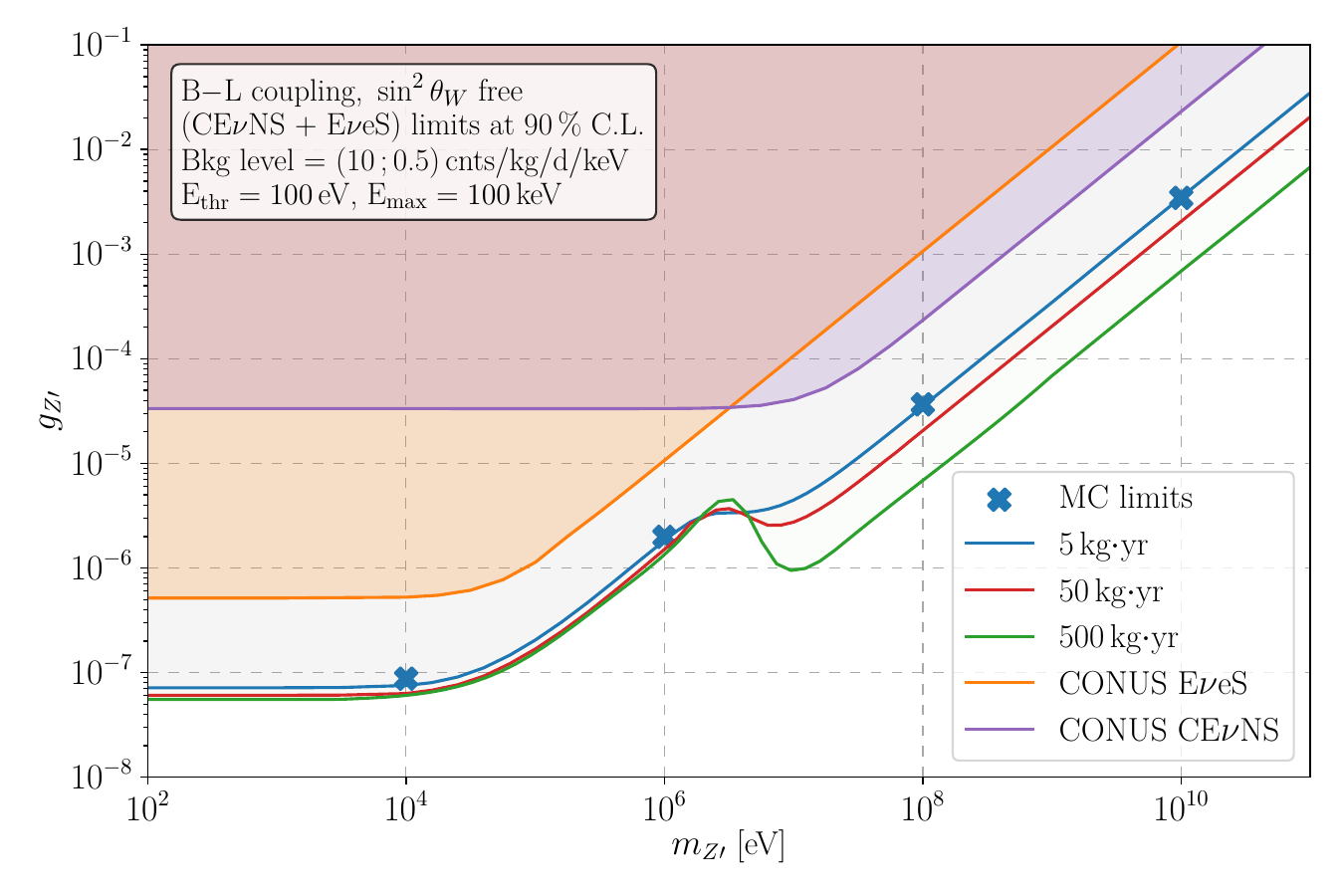}
    \caption{Expected exclusion limits at 90 \% C.L. on a light gauged B$-$L vector mediator for an extended energy region ($E_{I}\leq100$\,keV) and different exposures and detector thresholds. Background levels and detector thresholds are given in the boxes, while the exposures under study are illustrated with different colouring. Exemplary limits from MC-sampled data are represented by crosses to validate our results. For comparison \textsc{Conus} limits from their first BSM investigation are indicated in orange (E$\nu$eS-only) and purple (CE$\nu$NS-only)~\cite{CONUS:2021dwh}. In contrast to the previous case, the U(1)$_{\rm B-L}$ does not exhibit any degeneracy region.}
    \label{fig:light_vector_limits_b_l_long}
\end{figure}

%%%%%%%%%%%%%%%%%%%%%%%%%%%%%%%%%%%%%%%%%%%%%%%%%%%%%%%%%%%%%%%%%%%%%%%%%%%%%%%%%%%%%%%%%%%%%%%%%%%%

\begin{table}[H]
\resizebox{\textwidth}{!}{
    \centering
    \begin{tabular}{c|ccc}\hline
        Exposure / $E_{\mathrm{thr}}$ & 180~eV & 150~eV & 100~eV \\ \hline
        5~kg$\cdot$yr   & $8.0\cdot10^{-8}$ / $5.4\cdot10^{-4}$ & $7.7\cdot10^{-8}$ / $4.8\cdot10^{-4}$ & $7.2\cdot10^{-8}$ / $3.5\cdot10^{-4}$\\
        50~kg$\cdot$yr  & $6.3\cdot10^{-8}$ / $3.9\cdot10^{-4}$ & $6.7\cdot10^{-8}$ / $3.1\cdot10^{-4}$ & $6.1\cdot10^{-8}$ / $2.1\cdot10^{-4}$ \\
        500~kg$\cdot$yr & $5.7\cdot10^{-8}$ / $1.3\cdot10^{-4}$ & $5.6\cdot10^{-8}$ / $1.0\cdot10^{-4}$ & $5.5\cdot10^{-8}$ / $6.8\cdot10^{-5}$ \\ \hline
    \end{tabular}
    }
    \caption{Exemplary exclusion limits at 90\% C.L.\ for a light U(1)$_{\rm B-L}$ vector boson for an extended energy range ($E_{I}\leq100$\,keV). Values are given for regions where the experimental reach is dominated by E$\nu$eS ($m_{Z'}=1$\,keV) and CE$\nu$NS ($m_{Z'}=1$\,GeV), respectively.}
    \label{tab:light_vector_B_L_limits_long}
\end{table}

%%%%%%%%%%%%%%%%%%%%%%%%%%%%%%%%%%%%%%%%%%%%%%%%%%%%%%%%%%%%%%%%%%%%%%%%%%%%%%%%%%%%%%%%%%%%%%%%%%%%

\bibliography{literature} 
\bibliographystyle{JHEP}

%%%%%%%%%%%%%%%%%%%%%%%%%%%%%%%%%%%%%%%%%%%%%%%%%%%%%%%%%%%%%%%%%%%%%%%%%%%%%%%%

\end{document}